\documentclass[journal]{new-aiaa}
\usepackage[utf8]{inputenc}

\usepackage{bm}
\usepackage{amsmath}
\usepackage{color,soul}
\usepackage{todonotes}
\usepackage{colortbl}
\definecolor{myblue}{rgb}{0, 0.23, 0.64}
\definecolor{WVUblue}{rgb}{0, 0.16, 0.33}

\usepackage{hyperref}
\hypersetup{
    colorlinks=true,
    linkcolor=myblue,
    filecolor=magenta,
    citecolor=myblue,
    urlcolor=myblue,
}

\usepackage[flushleft]{threeparttable}

\usepackage{xcolor,pifont}
\usepackage{tablefootnote}
\usepackage{graphicx}
\usepackage[version=4]{mhchem}
\usepackage[group-separator={,}]{siunitx}
\usepackage{longtable,tabularx}
\usepackage{gensymb}
\usepackage{mathtools}
\usepackage{float}
\usepackage[ruled, vlined, linesnumbered]{algorithm2e}
\SetKwComment{Comment}{$\triangleright$\ }{}
\usepackage{subcaption}
\usepackage{mwe}
\usepackage{enumitem}
\usepackage{multirow}
\usepackage{booktabs}
\usepackage{lscape}
\usepackage{mdframed}
\usepackage{tabularx}
\usepackage{adjustbox}
\usepackage{dsfont}
\usepackage{xspace}
\usepackage{tcolorbox}
\usepackage[flushleft]{threeparttable}
\usepackage{comment}
\usepackage{enumitem}

\let\svthefootnote\thefootnote
\newcommand\freefootnote[1]{
  \let\thefootnote\relax
  \footnotetext{#1}
  \let\thefootnote\svthefootnote
}

\title{Encounter Geometry Effects on Space-Based Laser Debris Remediation and Estimation}

\author{Matthew C. Fox\footnote{Ph.D. Student, Department of Mechanical, Materials and Aerospace Engineering, Student Member AIAA.}, Gavin M. Baker\footnote{Ph.D. Student, Department of Mechanical, Materials and Aerospace Engineering, Student Member AIAA.}, David O. Williams Rogers\footnote{Ph.D. Student, Department of Mechanical, Materials and Aerospace Engineering, Student Member AIAA.}, and Hang Woon Lee\footnote{Assistant Professor, Department of Mechanical, Materials and Aerospace Engineering; hangwoon.lee@mail.wvu.edu. Member AIAA (Corresponding Author).}}
\affil{West Virginia University, Morgantown, WV, 26506}

\begin{document}

\freefootnote{This paper is a substantially revised version of the paper AIAA 2025-0980, presented at the 2025 AIAA SciTech Forum, Orlando, FL, January 6-10, 2025. It offers new results, an additional solution methodology, and a better description of the materials.}

\maketitle

\begin{abstract}
The escalating accumulation of orbital debris poses a critical threat to future space operations. Space-based lasers leveraging laser ablation have emerged as a promising approach for mitigating debris proliferation and preserving the orbital environment. Current literature, however, treats space-based laser debris remediation as a deterministic problem, assuming that momentum transfer and the resulting debris perturbations are precisely known. In reality, laser-to-debris engagement outcomes are inherently stochastic due to partially known debris characteristics. Compounding this challenge, estimating critical laser-matter parameters in situ, such as the momentum coupling coefficient, requires ablation that consequently perturbs the debris trajectory. This establishes a coupled ablation-and-estimation problem in which the laser platform and target debris encounter geometry influence remediation effectiveness and estimation accuracy. To address this problem, we present a joint ablation-and-estimation methodology that provides insights into the driving factors that make different encounter geometries improve or degrade overall remediation and estimation performance. Results across multiple coplanar and out-of-plane encounter geometries demonstrate how periapsis-lowering capacity, linear system observability, and nonlinear estimation performance evolve as laser parameters and relative orbit geometry vary. By identifying the key drivers behind these metrics, this study highlights critical considerations for the safe and effective operation of space-based lasers under uncertainty.
\end{abstract}

\section*{Nomenclature}

{\renewcommand\arraystretch{1.0}
\noindent\begin{longtable*}{@{}l @{\quad=\quad} l@{}}
$\bm{a}_\text{L2D}$ & laser-to-debris perturbing acceleration vector, \si{km~s^{-2}} \\
$\bm{A}$ & nonlinear dynamics Jacobian \\
$B$ & material-dependent coefficient relating $I$ to pulse duration $\tau$, \si{W~s^{1/2}/m^2} \\
$c_\text{diff}$ & diffraction constant \\
$c_\text{m}$ & momentum coupling coefficient, \si{N/MW} \\
$D$ & primary mirror diameter, \si{m} \\
$E$ & laser pulse energy, \si{J}\\
$\bm{G}$ & area matrix, \si{m^2} \\
$\bm{H}$ & nonlinear measurements Jacobian \\
$I$ & laser intensity, \si{W/m^2} \\
$\bar{I}$ & time-averaged laser intensity, \si{W/m^2} \\
$I_\text{p}$ & laser intensity needed for plasma ignition, \si{W/m^2} \\
$\bm{k}$ & relative range vector of debris with respect to a laser platform, \si{km} \\
$\bm{K}$ & Kalman gain \\
$L_\text{max}$ & maximum ablation distance, \si{km} \\
$L_\text{min}$ & minimum ablation distance, \si{km} \\
$\text{L2D}$ & laser-to-debris \\
$m_\text{D}$ & debris mass, \si{kg} \\
$M$ & laser beam quality factor \\
$\mathcal{N}$ & multivariate Gaussian distribution \\
$\bm{P}$ & state covariance matrix  \\
$\bm{q}$ & state vector \\
$\bm{Q}$ & dynamical system covariance matrix \\
$\bm{R}$ & measurement covariance matrix \\
$\bm{r}_\text{D}$ & debris position vector with respect to the center of Earth, \si{km}\\
$\bm{r}_\text{LP}$ & laser platform position vector with respect to the center of Earth, \si{km}\\
$\text{RMSE}$ & root mean square error\\
$T_{\text{eff}}$ & total electro-optical efficiency \\
$V$ & total number of sampled measurements \\
$\bm{W}$ & observability Gramian \\
$\alpha$ & azimuth, \si{deg} \\
$\beta$ & laser beam illumination angle, \si{deg} \\
$\gamma_\text{D}$ & debris radius, \si{m}\\
$\delta$ & elevation, \si{deg}\\
$\bm{\eta}(t_i)$ & measurement evaluated at discrete time step $t_i$ \\
$\lambda$ & laser wavelength, \si{nm} \\
$\mu_\text{E}$ & Earth's standard gravitational parameter, \si{km^3/s^2} \\
$\upsilon$ & pulse repetition frequency, \si{Hz} \\
$\bm{\nu}$ & measurement noise \\
$\tau$ & laser pulse duration, \si{s} \\
$\bm{\Phi}$ & state transition matrix \\
$\Psi$ & laser fluence, \si{kJ/m^2} \\
$\omega$ & dynamical system process noise \\
\end{longtable*}}

\section{Introduction}

The ever-growing space economy and the proliferation of low Earth orbit (LEO) spacecraft are closely linked to a rapid increase in the orbital debris population in LEO. Until recently, many space missions did not adequately account for a future in which large derelict spacecraft could contribute to the creation of smaller yet lethal orbital debris that jeopardizes the safe operation of spacecraft in LEO. The volume of key orbital bands in LEO is a finite resource, and failing to address the issue of orbital debris will inevitably render vital volumes of LEO inaccessible in the long term. To ensure a sustainable future for space operations, mitigation protocols such as the European Space Agency's Zero Debris approach \cite{ESA_zero_deb} aim to limit the production of orbital debris, and active debris removal (ADR) constitutes an immediate and necessary action to remediate existing orbital debris \cite{debris_whitehouse,NASA_ODMSP_2019}.

This looming threat to the long-term sustainability of space-based operations in LEO warrants the exploration of various ADR methodologies, where space-based lasers have been identified as a scalable and cost-effective solution to reducing the risks posed by orbital debris \cite{colvin2023cost}. Specifically, space-based laser remediation relies on either photon pressure to conduct just-in-time collision avoidance \cite{mason_asr_2011}, or laser ablation to deorbit debris, in addition to performing collision avoidance \cite{williams_asr_2025}. Photon pressure transfers momentum onto debris through absorption and reflection \cite{WALKER20232786}; however, the magnitude of the orbit modification is small compared to ablation-based systems operating at a similar power \cite{phipps2014Ladroit}. Conversely, laser ablation uses a sufficiently high-fluence laser beam to melt, vaporize, and ionize the surface of an object, resulting in an ablation plume that transfers momentum on targeted debris through a recoil force \cite{phipps_JSR_LAP_review_2010,phipps_bonnal_2016}. Within this mechanism, pulsed laser ablation has been shown to transfer higher momentum than continuous-wave ablation \cite{phipps2014Ladroit}.  Furthermore, the key figure of merit in laser ablation is the \textit{momentum coupling coefficient}, which measures the ratio of optical energy to transferred momentum.

Motivated by the benefits of space-based lasers, the literature has conducted extensive research on their use for debris remediation. For example, Ref.~\cite{SOULARD2014192} proposes a series of laser system designs to remediate small debris only (\textit{i.e.,} objects with characteristic length between 1 and \SI{10}{cm}). Similarly, Refs.~\cite{phipps2014Ladroit,phipps_bonnal_2016} propose the design of a laser system capable of nudging large debris (\textit{i.e.,} objects with a characteristic length greater than \SI{10}{cm}), deorbiting small debris, and performing just-in-time collision avoidance. Adopting these laser systems, several works have provided optimization frameworks to further enhance the mission performance by addressing constellation configuration design and operations \cite{williams_asr_2025,williams_asc_2025,Rogers_Reconfig_LP_consts_2026}, as well as their autonomous laser-to-debris (L2D) engagement scheduling \cite{Baker2025ReinforcementASC}.

The state-of-the-art in space-based laser debris remediation neglects a key aspect of their operations; that is, deterministic analyses do not capture how unknown debris characteristics and laser-matter interactions may result in highly uncertain perturbed debris trajectories, potentially endangering active satellites. Orbital debris remediation aims to reduce the risks posed to active satellites, but a means to obtain accurate state estimates of ablated debris has yet to be presented alongside the existing deterministic methodologies. With respect to laser ablation, the key parameter governing the debris's perturbation through a laser-matter interaction is the momentum coupling coefficient; however, a deterministic L2D engagement assumes complete knowledge of the debris's inertial and material characteristics, which is seldom true in reality. This governing parameter only appears in the debris's equations of motion during an L2D engagement, prompting an \textit{ablation-estimation} conundrum; consequently, the laser platform must \textit{ablate to estimate} the momentum coupling coefficient. Additionally, the sheer diversity of the debris field presents an equally diverse set of L2D engagements requiring a methodology to estimate ablated debris parameters and states that is also applicable to the wide variety of possible L2D engagements.

\begin{figure}[!h]
    \centering
    \includegraphics[width=\linewidth]{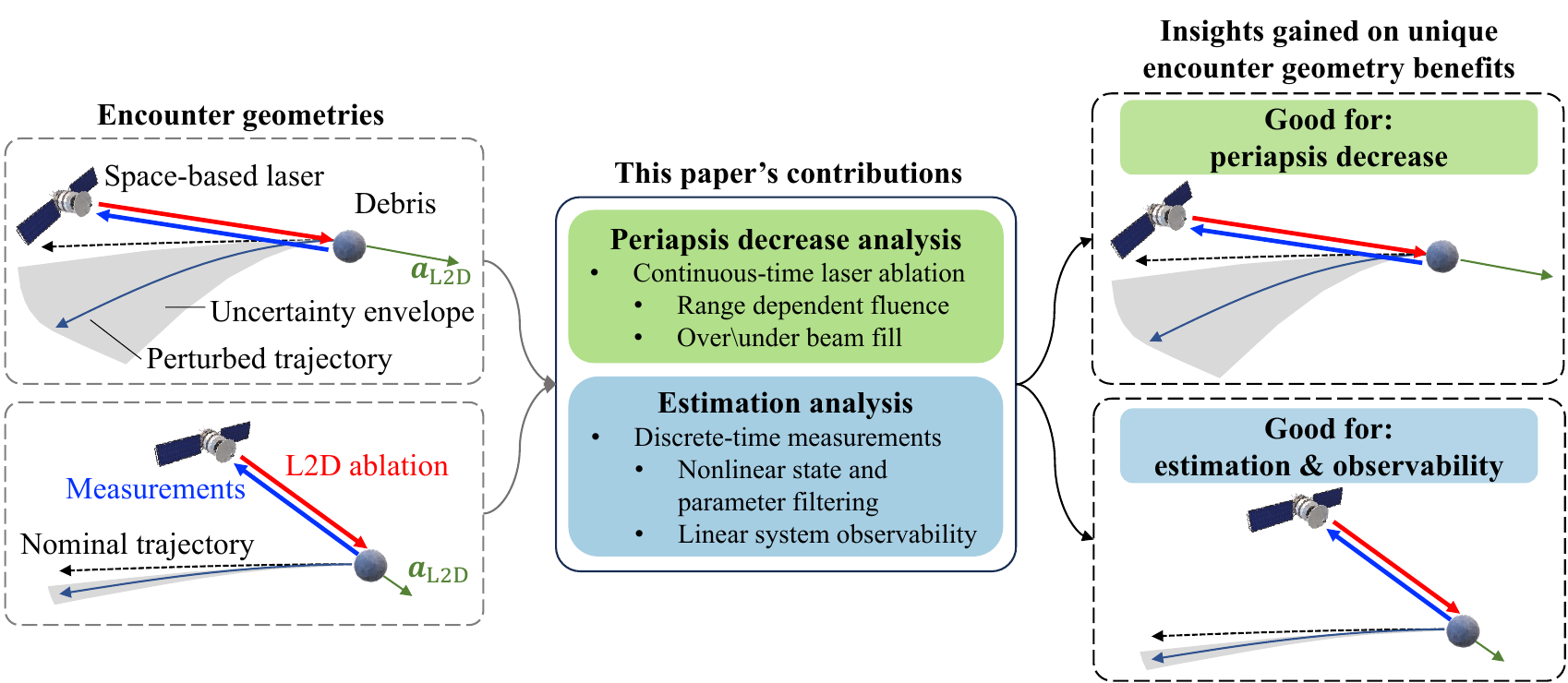}
    \caption{This paper characterizes an array of encounter geometries according to their state estimation, observability, and the amount of periapsis radius decrease on ablated debris.}
    \label{fig:overarching}
\end{figure}

For a given laser platform, orbital debris may take many different relative positions where an L2D engagement may occur, which this work defines as \textit{encounter geometries}. In light of the intrinsic uncertainty associated with space-based laser debris remediation operations, the myriad of possible L2D engagements offer various encounter geometries with the potential to provide safe, yet effective, L2D engagements. Figure~\ref{fig:overarching} provides an illustration of how two generic encounter geometries may be ideal for either lowering a debris's periapsis radius at the cost of an uncertain perturbed trajectory, or being ideal for state and parameter estimation at the cost of the debris's periapsis radius decrease when prioritizing highly accurate perturbed trajectories. The goal of remediating debris relies on significantly reducing the debris's periapsis radius; however, achieving this goal in a stochastic setting must not come to the detriment of estimated debris states. Each L2D engagement is likely to be different from another in reality due to the debris field's diversity and potential encounter geometries from which to engage the debris, which prompts the question: What are the driving factors that make a given encounter geometry good for prioritizing debris periapsis radius decrease versus estimation performance, or are they two sides of the same coin?

In response to this research question, this paper presents a joint remediation-and-estimation methodology used to determine clusters of encounter geometries that provide exceptional total debris periapsis radius reduction and low-magnitude estimation errors with respect to a diverse set of possible L2D engagements. The methodology presented in this work employs a hybrid Extended Kalman Filter (EKF) to estimate the momentum coupling coefficient and ablated debris states during an L2D engagement. Linear system observability is used as a means to assess the value of particular encounter geometries in contributing information to the nonlinear filter's updates over time. Additionally, the change in the ablated debris's periapsis radius over the course of an L2D engagement is computed to determine the remediation capacity of a given L2D engagement. These aspects of orbital debris remediation and estimation via space-based lasers are analyzed to identify driving factors across several representative metrics capturing the overall remediation and estimation performance of an L2D engagement. From these analyses, we identify common conditions that contribute to well-informed L2D engagements that also exhibit effective remediation capacities for coplanar and out-of-plane encounter geometries while also investigating the effects of different laser system designs. This paper is an extension of the authors' previously presented paper given in Ref.~\cite{Fox_SBL_perf_2025}.
            
The remainder of this paper is organized as follows: Section~\ref{sec:Preliminaries} provides a description of the technical content required to understand the L2D engagement dynamics and analysis employed in each numerical experiment. Section~\ref{sec:NumExp_comm_para} details the common parameters and assumptions shared across the three numerical experiments conducted in this work. Sections~\ref{sec:num_exp_1_cop} and~\ref{sec:num_exp_2_oop} describe the experimental setup and results obtained from analyses of coplanar and out-of-plane L2D engagements, respectively. Section~\ref{sec:num_exp_3_param} provides the experimental setup and results from a parametric analysis of laser parameters for small-, medium-, and large-power laser platforms engaging a single target debris. Sec.~\ref{sec:rec_lim} provides recommendations based on the obtained results to be considered with the limitations of the numerical experiments. Finally, Sec.~\ref{sec:Con} gives concluding remarks and suggests routes for future research based on insights gained from this work.

\section{Preliminaries} \label{sec:Preliminaries}

This section provides the necessary preliminary technical details to conduct the numerical experiments and reproduce the results presented in this work. In particular, this section is organized as follows: first, Sec.~\ref{subsec:Pulsed_Laser_Ablation} describes the mechanism of pulsed laser ablation and explains how the laser--matter interaction at the target debris surface results in a perturbing acceleration. Second, Sec.~\ref{subsec:Dynamical_System} presents a continuous-time dynamical system describing the motion of orbital debris and the laser platform during scenarios in which an L2D engagement is either active or inactive. Third, Sec.~\ref{subsec:Linear_System_Obs} details the formulation of linear system observability based on the nonlinear dynamics and measurements considered in this work. Lastly, Sec.~\ref{subsec:nonlinear_filter_EKF} describes the hybrid nonlinear filter used to perform state and parameter estimation, considering a continuous-time dynamical system and a discrete-time measurement model.

\subsection{Space-Based Pulsed Laser Ablation} \label{subsec:Pulsed_Laser_Ablation}

Pulsed laser ablation is the process of removing surface material from an object through high-intensity pulses of laser irradiation. The interaction of the debris's surface material with high-intensity laser pulses generates a plasma plume that expands perpendicular to the surface, ultimately changing the debris's trajectory. Previous work by Ref.~\cite{liedahl_2013_pulsed_shape} details the derivation of a closed-form perturbing acceleration due to this laser-matter interaction, where the resulting perturbing acceleration due to laser ablation is given in Eq.~\eqref{eq:Liedahl_LA_gen} \cite{liedahl_2013_pulsed_shape}.
\begin{subequations}
    \begin{align}
        \bm{a}_{\text{L2D}} &= \frac{c_{\text{m}} \bar{I} \bm{G} \cdot \hat{\bm{k}}}{m_\text{D}} \label{eq:Liedahl_LA_gen} \\
        \bm{k} &= \bm{r}_\text{D}-\bm{r}_\text{LP} = \left[k_x, k_y, k_z\right]^\text{T} \label{eq:range} \\
        \hat{\bm{k}} &= \frac{\bm{k}}{\left\lVert \bm{k} \right\rVert} \label{eq:laser_dir}
    \end{align}    
\end{subequations}
In Eq.~\eqref{eq:Liedahl_LA_gen}, the L2D perturbing acceleration, $\bm{a}_{\text{L2D}}$, is given as a function of variables related to the laser-matter interaction, laser parameters, and debris parameters. The momentum coupling coefficient, $c_\text{m}$, is a material-dependent coefficient capturing the coupling between delivered laser fluence and transferred momentum during the laser-matter interaction. The laser parameter, $\bar{I}$, denotes the time-averaged laser intensity, while $m_\text{D}$ and $\bm{G}$ denote the debris mass and area matrix parameters, respectively. Equation~\eqref{eq:range} describes the relative range, $\bm{k}$, of the target debris with respect to the laser platform, where $\bm{r}_{\text{D}}$, $\bm{r}_{\text{LP}}$, and $k_{(\cdot)}$ are the position vectors of the target debris, laser platform, and component-wise relative range in the Earth-Centered Inertial (ECI) reference frame, respectively. Additionally, the laser beam propagation unit vector, $\hat{\bm{k}}$, is defined in Eq.~\eqref{eq:laser_dir}. In this formulation, the target debris is assumed to have a constant mass during laser ablation, where the spot size of the laser beam delivered to the target debris is assumed to always overfill the target debris.

Equation~\eqref{eq:Liedahl_LA_gen} can be decomposed into terms that are variable with respect to the distance between the target debris and the laser platform. The time-averaged laser intensity, $\bar{I}$, can be expressed in terms of the laser fluence, $\Psi$, and the pulse repetition frequency, $\upsilon$, where the laser fluence is related to the laser intensity and pulse duration, $\tau$.
\begin{subequations}
    \begin{align}
        \bar{I} &= \Psi \upsilon \label{eq:time_avg_ints} \\
        \Psi &= I \tau \label{eq:fluence_ints}
    \end{align}
\end{subequations}
Using Eqs.~\eqref{eq:time_avg_ints} and \eqref{eq:fluence_ints}, the perturbing accelerations given in Eq.~\eqref{eq:Liedahl_LA_gen} can then be defined in terms of the laser fluence rather than the laser intensity. In addition to including range-dependent terms, the laser beam under or overfill can be expressed through a special formulation of the area matrix, $\bm{G}$ \cite{liedahl_2013_pulsed_shape}. Appendix~A details the manipulation of the L2D perturbing acceleration needed to account for these two factors, assuming that the target debris is spherical, where the final form of $\bm{a}_{\text{L2D}}$ expressed in the ECI frame is given as:
\begin{equation}
    \begin{bmatrix}
        a_{\text{L2D}_x}\\
        a_{\text{L2D}_y}\\
        a_{\text{L2D}_z}
    \end{bmatrix} = \begin{bmatrix}
        \frac{4 c_{\text{m}} E D^2 T_{\text{eff}} \upsilon \gamma_\text{D}^2}{c_\text{diff}^2 M^4 \lambda^2 \left\lVert \bm{k} \right\rVert^2 m_\text{D}} \left(\frac{2-3\cos(\beta)+\cos^3(\beta)}{3}\right) \\
        \frac{4 c_{\text{m}} E D^2 T_{\text{eff}} \upsilon \gamma_\text{D}^2}{c_\text{diff}^2 M^4 \lambda^2 \left\lVert \bm{k} \right\rVert^2 m_\text{D}} \left(\frac{2-3\cos(\beta)+\cos^3(\beta)}{3}\right) \\
        \frac{8 c_{\text{m}} E D^2 T_{\text{eff}} \upsilon \gamma_\text{D}^2}{c_\text{diff}^2 M^4 \lambda^2 \left\lVert \bm{k} \right\rVert^2 m_\text{D}}  \left( \frac{1 - \cos^3(\beta)}{3} \right)
    \end{bmatrix} \label{eq:a_L2D_G_eval_main}
\end{equation}
Relaxing the spherical debris assumption would require a generalized formulation of $\bm{G}$, which ultimately alters the final expressions given in Eq.~\eqref{eq:a_L2D_G_eval_main}; however, these alterations to the L2D perturbing acceleration can be made without loss of generality in the implementation of the nonlinear estimation problem formulated in this work.

It is important to note that Eq.~\eqref{eq:a_L2D_G_eval_main} considers a range-dependent momentum coupling coefficient by accounting for the range-dependent fluence term, but $c_\text{m}$ is known to also vary as the temperature of the ablation plume plasma changes over time \cite{phipps2014Ladroit}. Capturing the temperature variation of the plasma forming in the ablation plume involves an in-depth analysis of highly nonlinear interactions at an atomic level~\cite{Phipps_c_m0}, which falls outside the scope of this work; therefore, this work assumes that $c_{\text{m}}$ is a constant parameter rather than a variable due to variations in the ablation plume plasma temperature.

\subsection{Dynamical System} \label{subsec:Dynamical_System}

The dynamical system governing the motion of the laser platform over time assumes an acceleration due to the two-body central force with the Earth acting as the primary body. This dynamical system is also used for the target debris while not being ablated; conversely, the L2D perturbing acceleration due to laser ablation is added to the target debris's equations of motion during L2D engagements. Additionally, perturbing accelerations are not included in this dynamical system to provide insights into the impact of the L2D perturbing accelerations alone on all L2D engagements analyzed in this work. The equation of motion of the target debris while being ablated is given in Eq.~\eqref{eq:EOM_LA_vec},
\begin{equation}
    \centering
        \ddot{\bm{r}}_\text{D} = -\frac{\mu_\text{E}}{\left\lVert\bm{r}_\text{D}\right\rVert^3}\bm{r}_\text{D} + \bm{a}_\text{L2D} \label{eq:EOM_LA_vec}
\end{equation}
where $\mu_\text{E}$ is the standard gravitational parameter for Earth.

\subsection{Linear System Observability} \label{subsec:Linear_System_Obs}
Linear system observability is a widely adopted means to understand the quality of sampled measurements over time through various metrics obtained from the observability Gramian. In this work, the observability Gramian is used to determine the quality of information gained from sampled measurements across all state variables per time step over a full L2D engagement. The continuous time-varying dynamics and discrete-time measurement models used in this work are nonlinear, as shown in Eqs.~\eqref{eq:obs_dyn_eq} and \eqref{eq:obs_meas_eq}, respectively,
\begin{subequations}
    \begin{align}
        \dot{\bm{q}}(t) &= \bm{f}(\bm{q}(t)) \label{eq:obs_dyn_eq} \\
        \bm{\eta}(t_i) &= \bm{h}(\bm{q}(t_i)) \label{eq:obs_meas_eq}
    \end{align}
\end{subequations}
where the state vector, $\bm{q}(t)$, is used in the evaluation of the continuous-time nonlinear dynamics, $\bm{f}(\bm{q}(t))$, to propagate the state vector forward in time. The nonlinear measurement function, $\bm{h}(\bm{q}(t_i))$, captures the measurements, $\bm{\eta}(t_i)$, based on the state vector evaluated at a discrete time step, $t_i$.

A linearized system is created from the nonlinear dynamics, where linearizing Eqs.~\eqref{eq:obs_dyn_eq} and \eqref{eq:obs_meas_eq} requires the Jacobians of the nonlinear dynamics and measurement equations with respect to the system's state variables evaluated at a nominal trajectory. As the measurement equation is given in discrete time, the linear system observability is dependent on the ability to reconstruct the initial state vector at each discrete measurement time throughout the simulation time length. The proper reconstruction of a given initial state based on measurements sampled over time requires the observability Gramian, $\bm{W}(t_0,t_V)$, to be invertible. The observability Gramian is defined as follows:
\begin{equation}
    \centering
        \bm{W}(t_0,t_V) = \sum_{i=0}^{V}\bm{\Phi}(t_i,t_0)^\text{T} \bm{H}(t_i)^\text{T} \bm{H}(t_i)\bm{\Phi}(t_i,t_0) \label{eq:obs_gram_hybrid}
\end{equation}
where the initial time and total number of measurements are denoted as $t_0$ and $V$, respectively. The observability Gramian is also dependent on the state transition matrix, $\bm{\Phi}(t_i,t_0)$, and the measurement Jacobian, $\bm{H}(t_i)$.

Computing the observability Gramian over a single time step gives insight into how much information is gained about each state over a single time step, depending on variations of the state transition matrix and measurement Jacobian. The quality of a measurement interval can be obtained through a singular value decomposition of the observability Gramian through an inspection of the observability Gramian's singular values \cite{FRIEDMAN2018405_obs,Hinson_obs_diss_2014}.
        
\subsection{Nonlinear Filtering} \label{subsec:nonlinear_filter_EKF}

The EKF is used in this work to determine an estimate of the state space of the orbital debris targeted by the laser platform. The EKF assumes nonlinearity with a continuous-time system and discrete measurements. The nonlinear system dynamic model is given as:
\begin{equation*}
    \bm{\dot {q}}=\bm{f}(\bm{q},t) + \bm{\omega}, \qquad \bm{\omega} \thicksim \mathcal{N}(\bm{0},\bm{Q})
\end{equation*}
where $\bm{f}(\bm{q},t)$ is the governing nonlinear system dynamics function, and $\bm{\omega}$ denotes the zero-mean Gaussian process noise with a corresponding dynamical system covariance $\bm{Q}$. In both the dynamic model and measurement model, $\mathcal{N}$ indicates the multivariate Gaussian distribution. The state vector, $\bm{q}$, is defined in Eq.~\eqref{eq:q_EKF} as follows,
\begin{equation}
    \bm{q} = \left[x, y, z, \dot{x}, \dot{y}, \dot{z}, c_\text{m} \right]^\text{T} \label{eq:q_EKF}
\end{equation}
where the first six state variables are given as the position and velocity components of the target debris in a three-dimensional Cartesian system. In addition to the six state variables, $\bm{q}$ is augmented to include $c_\text{m}$, which allows parameter estimation to be performed. The time rate of change of $c_\text{m}$ is set to zero (\textit{i.e.}, $\dot{c}_\text{m}=0)$ as this parameter is assumed to be constant in this work. The measurement model vector $\bm{\eta}_i$ at time step $i$ is defined as \begin{equation*}
    \bm{\eta}_i = \bm{h}(\bm{q}_i)+\bm{\nu}_i, \qquad \bm{\nu}_i \thicksim \mathcal{N}(\bm{0},\bm{R}_i)
\end{equation*}
where $\bm{h}(\bm{q}_i)$ contains the measurement outputs that are to be disturbed by the zero-mean Gaussian error with covariance $\bm{R}_i$ and a measurement noise of $\bm{v}_i$. With equations of motion and measurements defined, the EKF is utilized to predict the next state vector by,
\begin{equation*}
    \dot{\hat{\bm{q}}} = \bm{f}(\bm{\hat{q}},t)
\end{equation*}
where the numerical integration of the nonlinear equations, $\dot{\hat{\bm{q}}}$, evaluated with respect to the estimated states provides the \textit{a priori} estimate of state vector $\bm{q}$. Within the same time step, the covariance matrix $\bm{P}$ is predicted. The evolution of the state covariance matrix is given as:
\begin{equation*}
    \dot{\bm{P}} = \bm{A}(t)\bm{P} + \bm{P}{\bm{A}(t)}^T+\bm{Q}
\end{equation*}
where the Jacobian of the nonlinear dynamics with respect to the state vector is given as follows,
\begin{equation*}
    \bm{A}(t) = \frac{\partial{\bm{f}}}{\partial{\bm{q}}}
\end{equation*}

After predictions are made, the initial estimates are then corrected and updated in the next step. This involves comparing the estimated state representation to the measured data that is available. If a measurement is available, the Kalman gain is computed as:
\begin{equation}
    \bm{K}_i = \bm{P}^-_i\bm{H}^T_i(\bm{H}_i\bm{P}^-_i\bm{H}^T_i + \bm{R}_i)^{-1} \label{eq:Kalman_gain}
\end{equation}
where the Jacobian of the measurement model is given as:
\begin{equation*}
    \bm{H}_i=\frac{\partial{\bm{h}}}{\partial{\bm{q}}}\bigg|_{\bm{\hat{q}}^-_i}
\end{equation*}
The state estimate and the covariance matrices are then updated. Finally, the \textit{a posteriori} estimates of both the state $\bm{\hat{q}}^+_i$ and the covariance $\bm{P}^+_i$ are given as:
\begin{equation*}
    \bm{\hat{q}}^+_i = \bm{\hat{q}}^-_i+\bm{K}_i \left[\bm{\eta}_i-\bm{h}_i(\bm{\hat{q}}^-_i)\right]
\end{equation*}
\begin{equation}
    \bm{P}^+_i = (\bm{I}-\bm{K}_i\bm{H}_i)\bm{P}^{-}_i \label{eq:cov_posteriori}
\end{equation}
where $\bm{I}$ is an identity matrix with conformable dimensions.

The root mean square error (RMSE) is a common metric that can be used to quantify the performance of an estimation algorithm. Considering the full set of sampled measurements, the target debris position and velocity RMSE values are calculated according to Eqs.~\eqref{eq:RMSE:pos} and \eqref{eq:RMSE:vel}, respectively.
\begin{subequations}
    \begin{align}
        \text{RMSE}_\text{pos} & = \sqrt{\frac{\sum_{i=1}^{V}\left\lVert\hat{\bm{q}}_i^{\text{pos}}-\bm{q}_i^{\text{pos}}\right\rVert^2}{V}} \label{eq:RMSE:pos} \\
        \text{RMSE}_\text{vel} & = \sqrt{\frac{\sum_{i=1}^{V}\left\lVert\hat{\bm{q}}_i^{\text{vel}}-\bm{q}_i^{\text{vel}}\right\rVert^2}{V}} \label{eq:RMSE:vel}
    \end{align}
\end{subequations}
In Eqs.~\eqref{eq:RMSE:pos} and \eqref{eq:RMSE:vel}, $(\cdot)^{\text{pos}}$ and $(\cdot)^{\text{vel}}$ denote truncated state vectors comprising solely position and velocity state elements, respectively.
        
\section{Numerical Experiments Assumptions and Common Parameters}\label{sec:NumExp_comm_para}

This work conducts three distinct numerical experiments to investigate the linear system observability, nonlinear estimation, and deorbiting performance for laser platform and target debris pairs across different encounter geometries. Each numerical experiment performs a unique laser platform and target debris pair initialization tailored to a specific design space, but similarities exist in the parameters needed to define the laser system, target debris, and nonlinear filter across each numerical experiment. In the nonlinear estimation analyses, the L2D perturbing acceleration is evaluated using the estimated debris states and $c_\text{m}$ values at each time step of a given L2D engagement. The relative range and angle measurement standard deviations used in this work are given in Refs.~\cite{Lee_LaserBased_RelNav} and \cite{HIPPELHEUSER2021_meas_std_dev}, respectively. These common laser system, target debris, and nonlinear filter parameters are provided in Table~\ref{table:common_para}.
\begin{table}[!h]
\caption{Common Numerical Experiment parameters \cite{phipps2014Ladroit,HIPPELHEUSER2021_meas_std_dev,Lee_LaserBased_RelNav}}
\centering
\begin{tabular}{lll}
\hline
\hline
Parameter & Value & Unit \\
\hline
\multicolumn{3}{c}{\underline{\textit{Laser system}}} \\
Pulse energy, $E$ & 380 & \si{J}\\
Fluence, $\Psi$ & 8.50 &  \si{kJ/m^2}\\
Pulse repetition frequency, $\upsilon$ & 56.0 & \si{Hz} \\
Time-averaged intensity, $\bar{I}$ & 476.0 & \si{kW/m^2} \\
Wavelength, $\lambda$ & 355.0 &  \si{nm}\\
Pulse duration, $\tau$ & 0.10 & \si{ns} \\
Primary mirror diameter, $D$ & 1.50 & \si{m} \\
Beam quality factor, $M^2$ & 2.00 & - \\ \hline
\multicolumn{3}{c}{\underline{\textit{Target debris}}}\\
Mass, $m_\text{D}$ & 4.7890 & \si{kg} \\
Density & 2710.0 & \si{kg/m^3} \\
Characteristic length & 15.0 & \si{cm} \\
Momentum coupling coefficient, $c_{\text{m}}$ & 99.0 & \si{N/MW} \\ \hline
\multicolumn{3}{c}{\underline{\textit{Nonlinear filter}}} \\
Propagation time step size & 1.0 & \si{s} \\
Measurement time step size & 1.0 & \si{s} \\
Position process noise standard deviation & 1.00 & \si{m} \\
Velocity process noise standard deviation & \num{5.00e-3} & \si{m/s} \\
Angle measurement noise standard deviation & 0.03 & \si{deg} \\
Range measurement noise standard deviation & \num{3.16e-3} & \si{m}\\
Initial $c_{\text{m}}$ estimate & 10.0 & \si{N/MW} \\
\hline
\hline
\label{table:common_para}
\end{tabular}
\end{table}

\subsection{L2D Engagement Constraints} \label{subsubsec:L2D_cons}

To initiate an L2D engagement, the target debris position relative to the laser platform must be less than and greater than the maximum and minimum ablation range, respectively. Once this range constraint is satisfied, a single time-step forward propagation is performed with laser ablation active. If laser ablation is expected to decrease the periapsis radius, then the L2D engagement begins; otherwise, the L2D engagement does not start. This periapsis radius decrease constraint is checked at each simulation time step. The L2D engagement concludes at the time step when either the target debris exits the allowable ablation range or when the expected periapsis radius change over a single time step becomes nonnegative. In addition to the ablation range constraints, an L2D engagement concludes if the ablated debris's altitude falls below \SI{100}{km}, where atmospheric drag is assumed to force debris reentry. For each numerical experiment, the simulation time length for a given laser platform and target debris pair begins and ends at the first and last time steps that the L2D engagement constraints are met, respectively.

\subsection{Measurement Model} \label{subsubsec:Meas_Model}
The measurement model used in this work consists of two angle measurements and a single range measurement made on a given target debris with respect to the laser platform. The range measurement is given by Eq.~\eqref{eq:range}, and the two angle measurements are the azimuth, $\alpha$, and elevation, $\delta$, angles given in Eqs.~\eqref{eq:meas_az} and \eqref{eq:meas_el}, respectively.
\begin{subequations}
    \begin{align}
        \alpha &= \arctan2 \left( k_y,k_x \right) \label{eq:meas_az} \\
        \delta &= \arcsin \left( \frac{k_z}{\sqrt{k_x^2+k_y^2+k_z^2}} \right)\label{eq:meas_el}
    \end{align}
\end{subequations}

\subsection{Assumptions}

Without loss of generality in the application of the proposed estimation methodology, several key assumptions are made in each numerical experiment. First, the target debris is considered to be a non-rotating, spherical, homogeneous solid aluminum rigid body. These simplifying assumptions are made such that all analyses focus on overall L2D engagement performance due to variations in encounter geometry, rather than changes in local L2D engagement performance induced by the increasingly complex debris characteristics. Relaxing these assumptions would require a reformulation of the L2D perturbing accelerations and accounting for rotational dynamics; however, the implementation in the overall nonlinear estimation framework would remain unchanged, as the state vector to be estimated can be augmented to include additional state variables as needed. Second, the laser-matter interaction assumes a negligible mass loss due to laser ablation. Finally, the laser platform and target debris are assumed to be initialized in circular orbits, and results and recommendations are given within the scope of encounter geometries based on initially circular debris orbits. This is a simplifying assumption made for each numerical experiment to reduce a degree of freedom in the debris's state initialization; however, the nonlinear estimation methodology presented in this work makes the initially circular debris orbit assumption without loss of generality to elliptic debris orbits, as elliptic encounter geometries occur following the first time step of laser ablation due to the L2D perturbing acceleration.

\section{Numerical Experiment 1: Coplanar L2D Engagement} \label{sec:num_exp_1_cop}

The goal of this numerical experiment is to investigate how the encounter geometry between a given laser platform and target debris in coplanar orbits impacts the target debris periapsis radius decrease, linear system observability, and nonlinear estimation performance. This section is organized as follows: Sec.~\ref{subsec:exp_set_1_cop} details the setup to initialize the numerical experiment; and Sec.~\ref{subsec:results_1_cop} presents the results for target debris periapsis radius decrease, linear system observability, and EKF performance for each target debris and laser platform pair analyzed.

\subsection{Experimental Setup} \label{subsec:exp_set_1_cop}

Multiple simulation cases are analyzed in this numerical experiment, where each case assumes the operation of a laser platform and target debris in coplanar orbits. For each case, a single laser platform is initialized on the $x$-axis, and a single target debris is initialized based on the L2D engagement constraints for the laser platform. This initialization is performed such that the laser platform executes an L2D engagement that lowers the target debris periapsis at the first simulated time step. The simulation is then propagated forward until at least one L2D engagement constraint is violated. The target debris initialization begins by finding a cone of visibility, accounting for Earth occultation with respect to the specific laser platform being initialized. Assuming that occultation due to the Earth occurs at a point where visibility is lost allows for lines to be drawn from the laser platform that extend tangent to a circle with a radius equal to the Earth's radius plus \SI{100}{km}. Initializing any circular target debris orbit with a radius larger than the Earth's radius allows for intersecting secant lines to be made along the visibility tangent lines that originate from the laser platform while intersecting the target debris orbit at two points on each secant line. These intersections create two solutions per secant line, and the solution with the largest distance from the laser platform is chosen for the initial target debris position. Cases are initialized with the target debris and laser platform having a positive angular momentum, so an appropriate choice in the target debris initialization must be made to ensure an L2D engagement is maintained following the first time step. If the target debris altitude is less than the laser platform altitude, then the secant intersection below the laser platform is chosen; on the other hand, the intersection at an altitude greater than that of the laser platform is used in cases where the target debris altitude is greater than the laser platform. The target debris and laser platform states are propagated forward in time until the L2D engagement constraints are met, and the final states in this propagation are used as the initial states in the L2D engagement analysis.

Twelve cases are initialized for this work, and relevant simulation parameters for each case are detailed in Table~\ref{table:num1_multicase}, where the cases are denoted with the first index being associated with a laser platform, and the second index denoting the debris initialized for the laser platform. Each case in this experiment is simulated as a single independent instance, where a single laser platform conducts an L2D engagement on a single targeted debris. Laser platform altitudes are varied from \SI{300}{km} to \SI{1650}{km} in \SI{450}{km} increments, while target debris altitudes are initialized based on the current laser platform's altitude. Target debris altitudes are created relative to the laser platform altitude, where the minimum and maximum target debris altitudes are set to \SI{75}{km} above and below the maximum laser ablation range, respectively. The target debris altitude is also stepped at \SI{170}{km} intervals starting from the first initialization for each respective laser platform. If the laser platform is initialized with a maximum ablation range extending into the radius of the Earth, then the first initialized target debris for that laser platform is set to \SI{120}{km}. Additionally, an encounter geometry and performance mapping is presented at the end of this section.
\begin{table}[!h]
    \caption{All coplanar case study initialization parameters.}
    \centering
    \begin{tabular}{lrrrrr}
        \hline
        \hline
        Case & Laser platform altitude, \si{km} & Target debris altitude, \si{km} & Simulation time, \si{s} \\
        \hline
        $(1, 1)$ & 300.0 & 120.0 & 109.0 \\
        $(1, 2)$ & 300.0 & 290.0 & 717.0 \\
        $(1, 3)$ & 300.0 & 460.0 & 1508.0 \\
        $(2, 1)$ & 750.0 & 517.22 & 910.0 \\
        $(2, 2)$ & 750.0 & 687.22 & 2218.0 \\
        $(2, 3)$ & 750.0 & 857.22 & 2030.0 \\
        $(3, 1)$ & 1200.0 & 927.22 & 970.0 \\
        $(3, 2)$ & 1200.0 & 1137.22 & 2230.0 \\
        $(3, 3)$ & 1200.0 & 1307.22 & 2090.0 \\       
        $(4, 1)$ & 1650.0 & 1417.22 & 1024.0 \\       
        $(4, 2)$ & 1650.0 & 1587.22 & 2205.0 \\       
        $(4, 3)$ & 1650.0 & 1757.22 & 2133.0 \\       
        \hline
        \hline
    \end{tabular}
    \label{table:num1_multicase}
\end{table}

\subsection{Experiment Results} \label{subsec:results_1_cop}

\subsubsection{Target Debris Periapsis Radius Decrease and Linear System Observability Analysis} \label{subsub:num1_peri_obs}

\begin{figure}[!h]
    \centering
    \includegraphics[width=\linewidth]{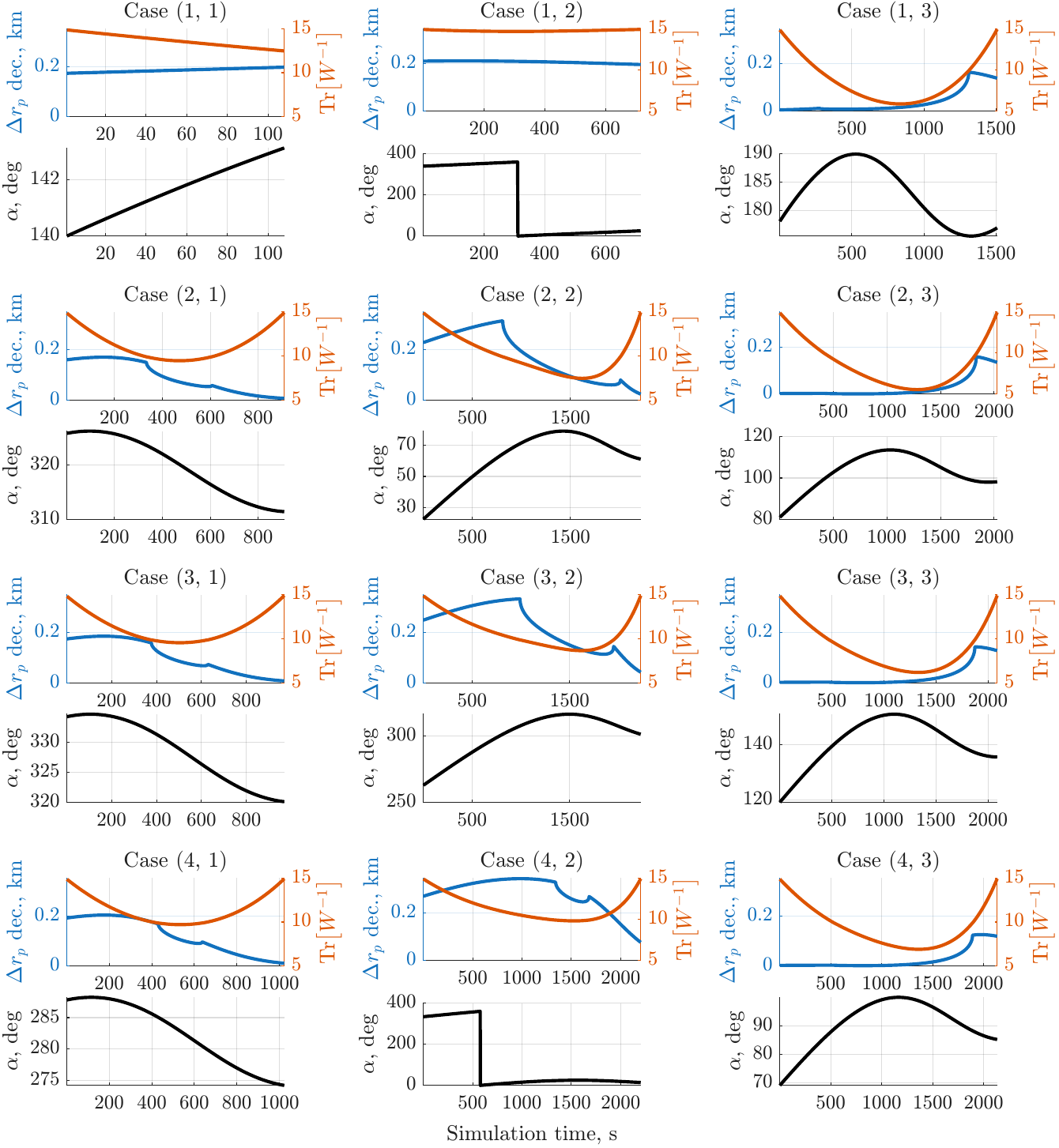}
    \caption{Periapsis radius decrease and observability Gramian inverse matrix trace variation over the simulation time length for each coplanar case.}
    \label{fig:num1_peri_obs_ALL}
\end{figure}

The total periapsis radius decrease magnitude of the target debris at the end of the L2D engagement for each coplanar case is given in Table~\ref{table:num1_peri_dec}, and the target debris periapsis radius decrease magnitude across each simulation time step for all coplanar cases is denoted as $\Delta r_\text{p}$ and shown in Fig.~\ref{fig:num1_peri_obs_ALL}. Case (3, 2) has the longest L2D engagement, whereas Case (4, 2) has the third longest simulation time with \SI{97}{s} fewer seconds of ablation delivered to its respective target debris while having reduced its respective target debris's periapsis radius by \SI{133.19}{km} more than Case (3 2); Case (4, 2)'s target debris and laser platform orbits and propagated states during the L2D engagement are shown in Fig.~\ref{fig:num1_42_traj}, where time steps with the maximum periapsis lowering decrease magnitude and minimum $\text{Tr}\left[W^{-1}\right]$ are visualized as well.
\begin{table}[!h]
    \caption{Total periapsis decrease magnitude and final position RMSE for all coplanar cases.}
    \centering
    \begin{tabular}{lrr}
        \hline
        \hline
        Case & Total $r_p$ decrease, \si{km} & $\text{RMSE}_{\text{pos}}$, \si{km} \\
        \hline
        $(1, 1)$ & 18.0103 & 0.0023 \\
        $(1, 2)$ & 146.5748 & 0.0013 \\
        $(1, 3)$ & 59.7390 & 0.0010 \\
        $(2, 1)$ & 81.4679 & 0.0016 \\
        $(2, 2)$ & 378.4832 & 0.0011 \\
        $(2, 3)$ & 53.9281 & 0.0010 \\
        $(3, 1)$ & 99.8046 & 0.0012 \\
        $(3, 2)$ & 484.5922 & 0.0011 \\
        $(3, 3)$ & 53.9554 & 0.0011 \\       
        $(4, 1)$ & 121.0197 & 0.0011 \\       
        $(4, 2)$ & 617.7815 & 0.0012 \\       
        $(4, 3)$ & 52.7843 & 0.0011 \\       
        \hline
        \hline
    \end{tabular}
    \label{table:num1_peri_dec}
\end{table}
\textbf{\begin{figure}[!h]
    \centering
    \includegraphics[width=.7\linewidth]{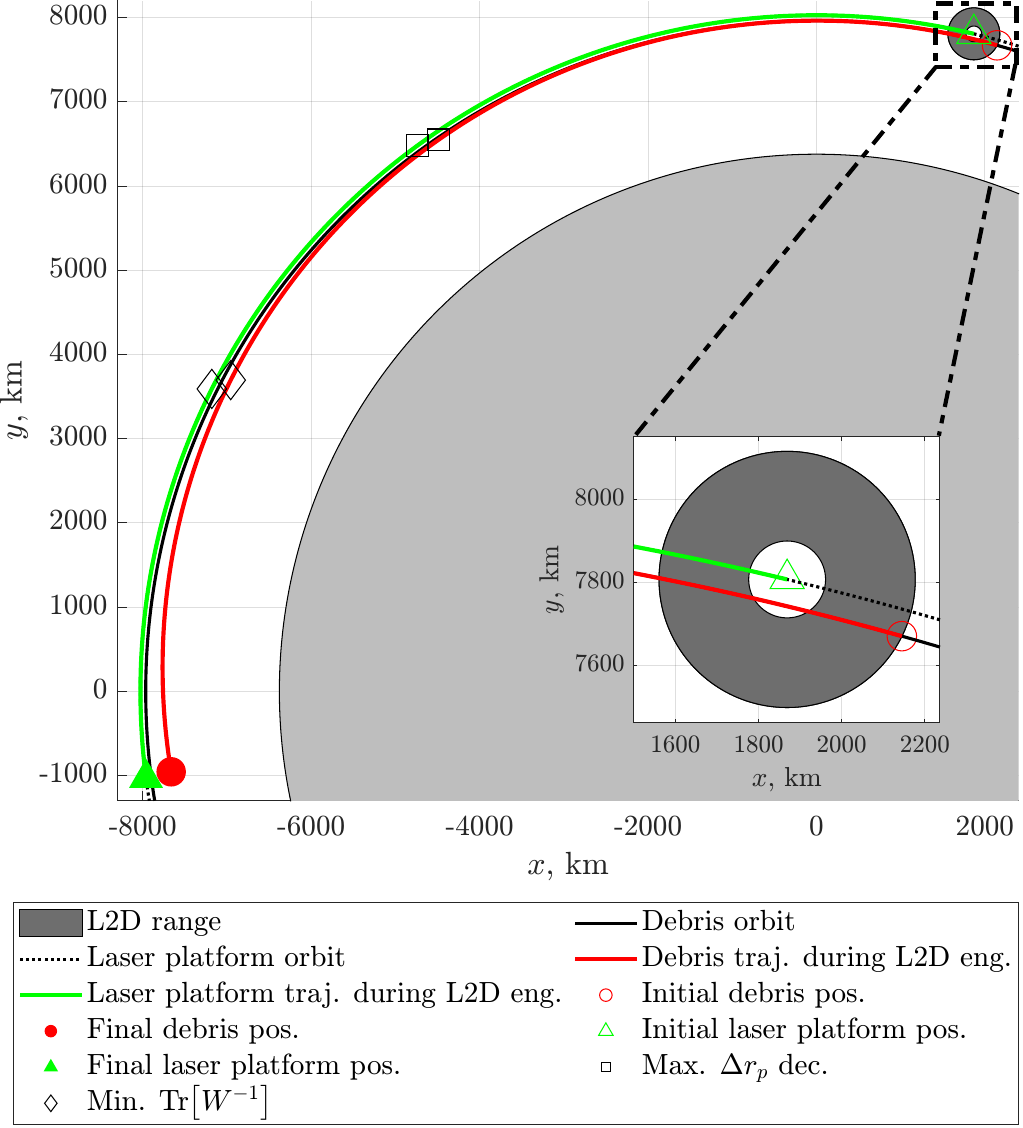}
    \caption{Coplanar Case (4, 2) target debris and laser platform orbits and propagated trajectories during the case's L2D engagement.}
    \label{fig:num1_42_traj}
\end{figure}}

Cases with a longer L2D engagement interval generally result in a greater total periapsis radius decrease, where the time interval for the L2D engagement increases for coplanar cases as the relative altitude between the laser platform and target debris decreases. In Fig.~\ref{fig:num1_peri_obs_ALL}, cases in the rightmost column have a target debris initialized with an altitude greater than the laser platform. In each of these cases, the expected positive periapsis decrease magnitude constraint is the last L2D engagement constraint satisfied before the debris can be ablated (as given in Sec.~\ref{subsubsec:L2D_cons}); consequently, the first time step of ablation for these cases results in a minimum periapsis decrease magnitude, where a maximum is approached towards the end of the simulation as the decrease in the relative range between the laser platform and target debris results in a larger magnitude laser fluence delivered to the target debris. For all other cases, the target debris is initialized with an altitude below the laser platform. In these cases, the L2D engagement begins as the range constraint is the final L2D engagement constraint to be satisfied before the debris can be ablated (as given in Sec.~\ref{subsubsec:L2D_cons}); therefore, the periapsis decrease magnitude quickly approaches a maximum as the relative range between the target debris and laser platform decreases, and the engagement ends with the periapsis radius decrease magnitude at a minimum, as the next time step would result in the periapsis lowering constraint being violated. The exceptions to these trends are Cases (1, 1) and (1, 2). Case (1, 1) has an L2D engagement that ended abruptly as the ablated debris's periapsis reduction allows the debris to be considered as deorbited. Case (1, 2) also does not follow the general trends in the periapsis reduction as the perturbed debris quickly exits the operational range of the laser platform during the L2D engagement.

The trace of the observability Gramian inverse across each time step of the simulation is given for each case in Fig.~\ref{fig:num1_peri_obs_ALL}. The magnitude of the observability Gramian elements depends on the dynamical system and measurement Jacobians, so the minimum $\text{Tr}\left[W^{-1}\right]$ magnitude occurs near time steps where the periapsis decrease magnitude and the measurement rate of change over a single time step are large.

\subsubsection{EKF Performance} \label{subsubsec:exp_1_ekf}

\begin{figure}[!h]
    \centering
    \includegraphics[width=\linewidth]{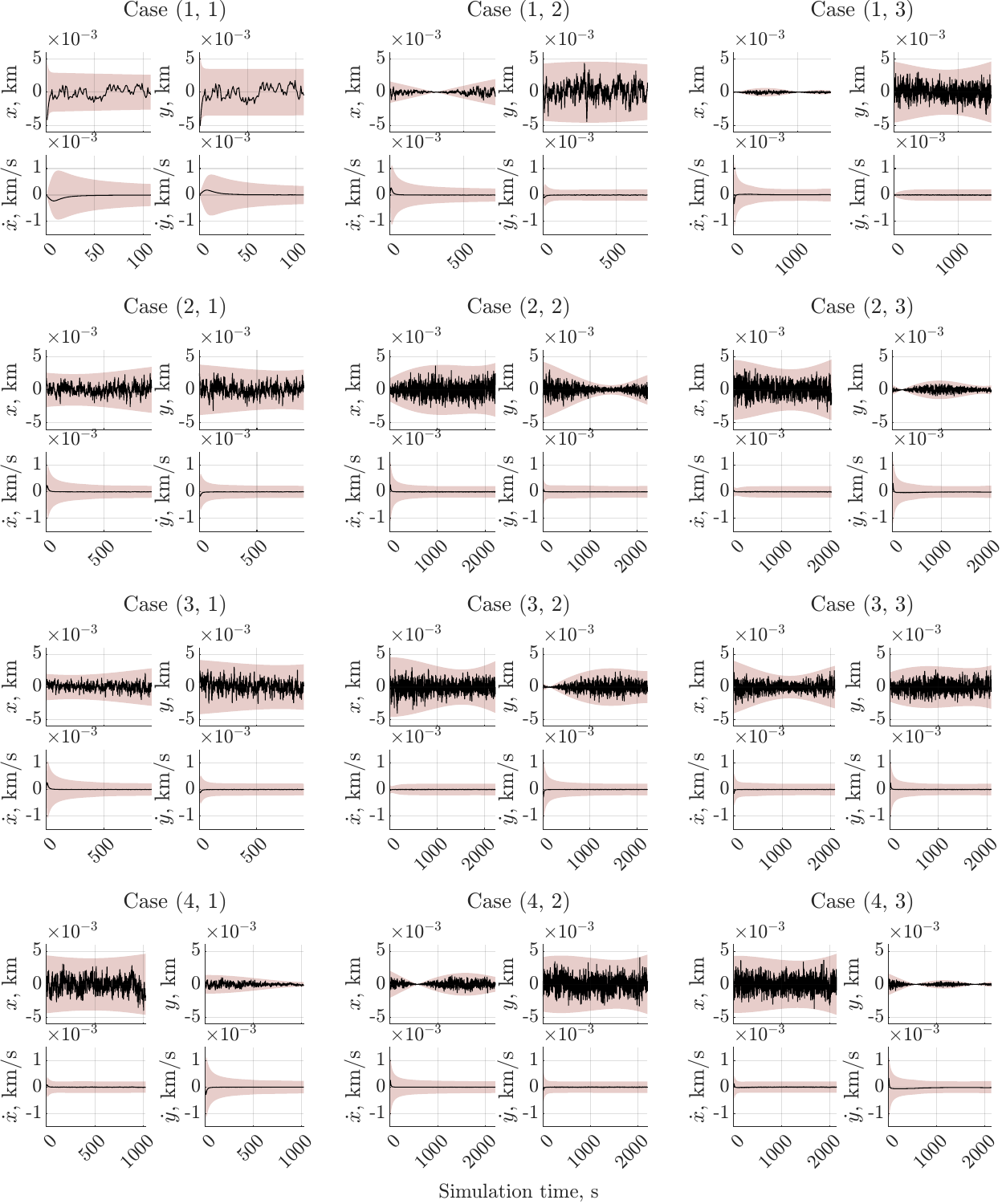}
    \caption{Target debris $x$ and $y$ position and velocity state errors (black) and corresponding three standard deviation bounds (red) over each coplanar case's simulation time length.}
    \label{fig:num1_state_err_ALL}
\end{figure}

The target debris position and velocity state errors over the simulation time length are shown in Fig.~\ref{fig:num1_state_err_ALL}, where the best-performing case with respect to the position RMSE is Case (1, 3). Cases with a longer L2D engagement are observed to have a lower position RMSE as compared to other cases, where the longer set of measurements during the L2D engagement contributes to lowering the RMSE over time. The magnitude of $\text{Tr}\left[W^{-1}\right]$ is seen to act as a predictor of the increased estimation performance, as cases that are more observable result in a lower position RMSE magnitude at the end of their respective L2D engagement.

\begin{figure}[!h]
    \centering
    \includegraphics[width=.7\linewidth]{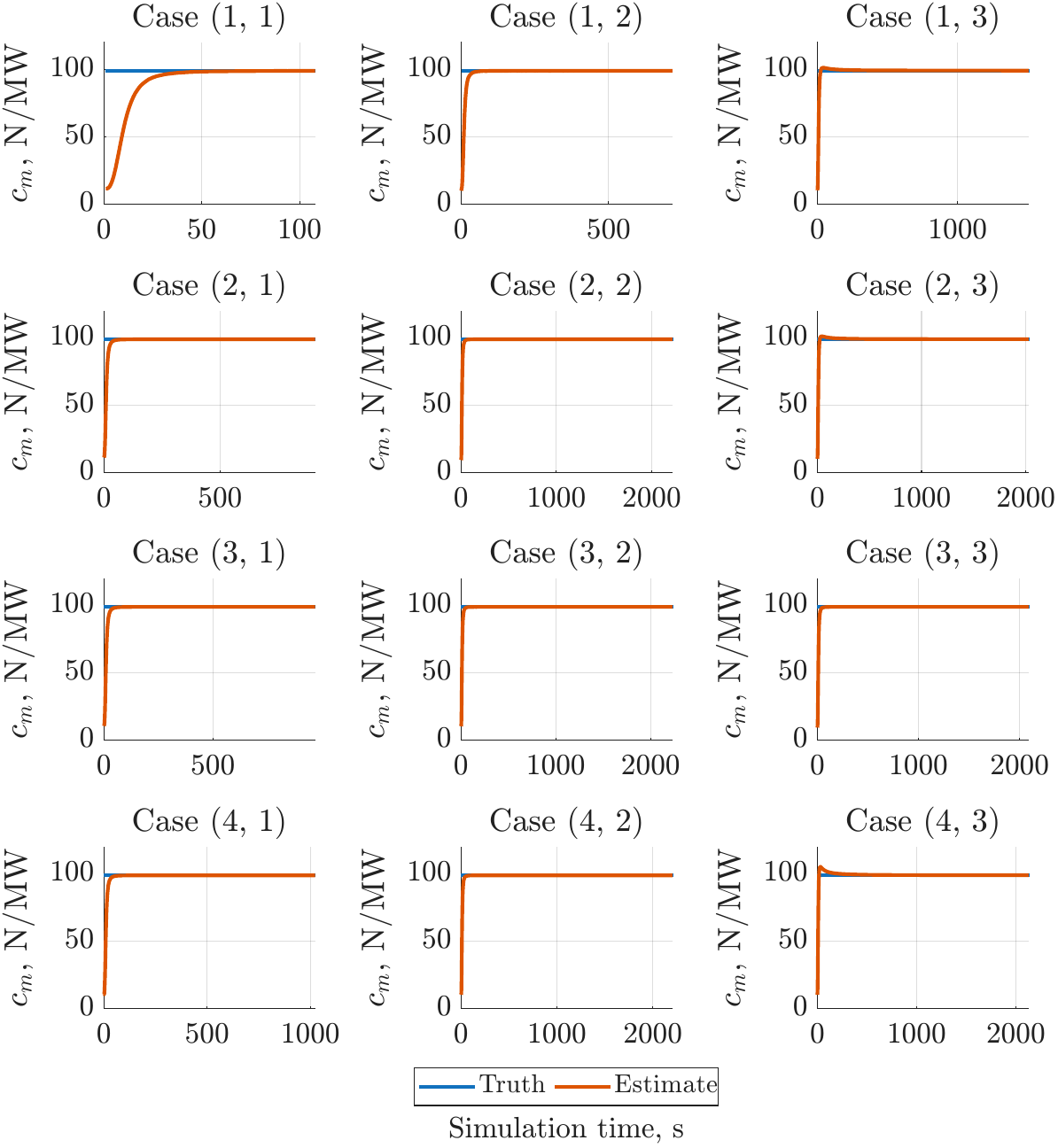}
    \caption{$c_\text{m}$ estimate and $c_\text{m}$ truth values over each coplanar case's simulation time length.}
    \label{fig:num1_cm_err_ALL}
\end{figure}

Performance of the $c_\text{m}$ estimation for each coplanar case is shown in Fig.~\ref{fig:num1_cm_err_ALL}, where all cases closely approach the $c_\text{m}$ truth value by the end of the L2D engagement. One would expect to see improved state estimation performance for L2D engagements where the target debris is not significantly perturbed from its nominal trajectory, but the coplanar analyses show that a larger magnitude perturbation has the counterintuitive result of decreasing the $\text{Tr}\left[W^{-1}\right]$ magnitude due to the role that the dynamical system Jacobian plays in the observability Gramian computation; thus, for coplanar L2D engagements, conducting an L2D engagement that significantly perturbs the debris will effectively decrease the error in estimated states.

\subsection{Encounter Geometry and L2D Engagement Performance Mapping}

\begin{figure}[!h]
    \centering
    \includegraphics[angle=0, width=0.55\linewidth]{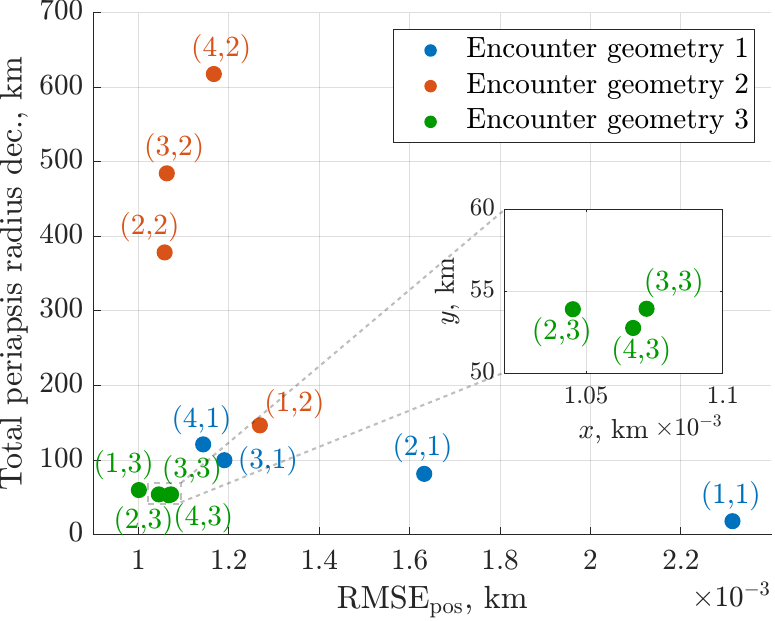}
    \caption{A mapping of the total periapsis radius reduction and final $\text{RMSE}_{\text{pos}}$ magnitudes of each case considered for all coplanar orbits.}
    \label{fig:Num1_geom_perf_map}
\end{figure}

Based on the debris's periapsis radius lowering, linear system observability, and nonlinear estimation performance analyses conducted in this experiment, a mapping of the trade-off between ablated debris's periapsis decrease and final $\text{RMSE}_\text{pos}$ is given in Fig.~\ref{fig:Num1_geom_perf_map}. This mapping shows how groupings of similar performance appear for debris that are initialized in similar encounter geometries with respect to each distinct laser platform considered in this experiment. Encounter geometry groupings are defined as 1, 2, and 3, which correspond to the encounter geometries where each considered laser platform performs an L2D engagement with their respective first, second, and third initialized debris. The natural grouping of encounter geometries across all four laser platforms analyzed in this experiment reinforces the overall dependency that a given L2D engagement's performance has on encounter geometry.

\section{Numerical Experiment 2: Out-of-Plane L2D Engagement} \label{sec:num_exp_2_oop}

The second numerical experiment for this work extends the methodology of the coplanar analysis performed in Sec.~\ref{sec:num_exp_1_cop} to consider out-of-plane L2D engagements. Section~\ref{subsec:exp_set_2_oop} details the experimental setup needed to analyze the more complex out-of-plane L2D engagements. Results of target debris periapsis radius decrease, linear system observability, and EKF performance for this numerical experiment are presented in Sec.~\ref{subsec:results_2_plane}.

\subsection{Experimental Setup} \label{subsec:exp_set_2_oop}

Investigating the performance of an L2D engagement when the laser platform and target debris do not exist in the same orbital plane requires a different initialization from the coplanar case. The laser platform's circular orbit, along with a specified maximum engagement range, defines a torus, where a toric section defining all possible L2D engagements can be created from the intersection of this torus with the target debris's orbital plane. The toric section can then be used to initialize multiple target debris across the various toric sections created by varying the target debris's semi-major axis and the relative inclination between the laser platform and target debris. A detailed description of the toric section initialization is given in Appendix~B. For this numerical experiment, 48 unique cases of different encounter geometries are simulated, where a single laser platform performs an L2D engagement on a single initialized target debris in each case. Considering each out-of-plane L2D engagement, a $k$-means clustering approach is used to determine two clusters of encounter geometries analyzed in this experiment. The encounter geometry clusters are created in a metric space spanning the total periapsis decrease magnitude, position RMSE, and the percent error of the $c_\text{m}$ estimate given at the final time step of a case's simulation time length. Clusters and their respective centroids are determined via MATLAB's \texttt{kmeans} clustering function \cite{MATLAB}, which is called with default parameters and an input of two clusters. To illustrate the characteristics of each cluster, one representative case from each cluster is selected for a detailed comparative analysis. A representative case is determined as a case in a cluster with the minimum Euclidean distance to its respective cluster centroid. The two representative cases are then subjected to the same periapsis radius decrease, linear system observability, and nonlinear estimation analyses presented throughout Sec.~\ref{sec:num_exp_1_cop}.

\subsection{Experiment Results} \label{subsec:results_2_plane}

This subsection presents a discussion of the two out-of-plane encounter geometry clusters and details the selection of the out-of-plane representative cases as well as their comparison across periapsis radius reduction and nonlinear estimation performance.

\subsubsection{Out-of-Plane Case Clustering}
\begin{figure}[!h]
    \centering
        \begin{subfigure}{.45\linewidth}
        \centering
        \includegraphics[width=\linewidth]{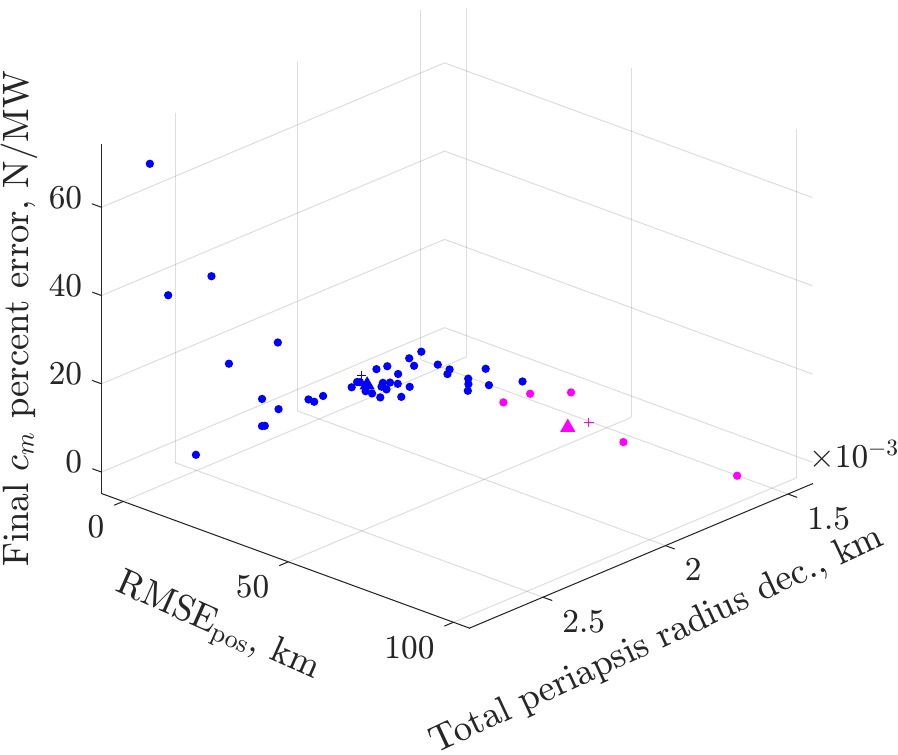}
        \caption{View of total periapsis radius decrease magnitude versus position RMSE clustering.}
        \label{fig:num2_metricClust_iso}
    \end{subfigure}
    \hfill
    \begin{subfigure}{.45\linewidth}
        \centering
        \includegraphics[width=\linewidth]{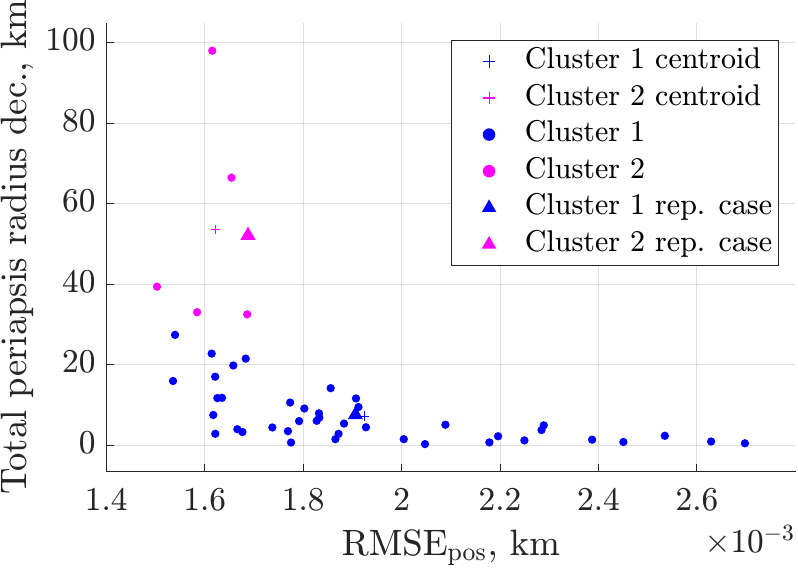}
        \caption{View of total periapsis radius decrease magnitude versus position RMSE clustering.}
        \label{fig:num2_metricClust_xy}
    \end{subfigure}
    \hfill
    \begin{subfigure}{.45\linewidth}
        \centering
        \includegraphics[width=\linewidth]{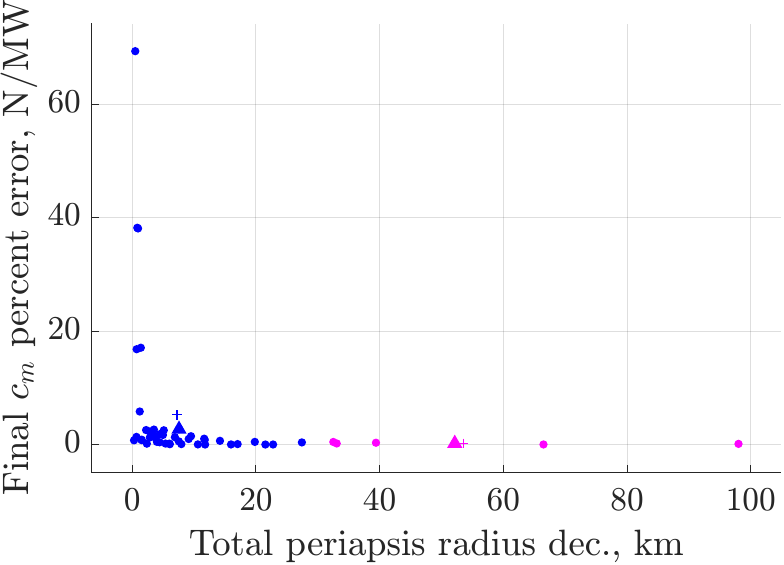}
        \caption{View of final $c_\text{m}$ percent error versus total periapsis radius decrease magnitude clustering.}
        \label{fig:num2_metricClust_yz}
    \end{subfigure}
    \hfill
    \begin{subfigure}{.45\linewidth}
        \centering
        \includegraphics[width=\linewidth]{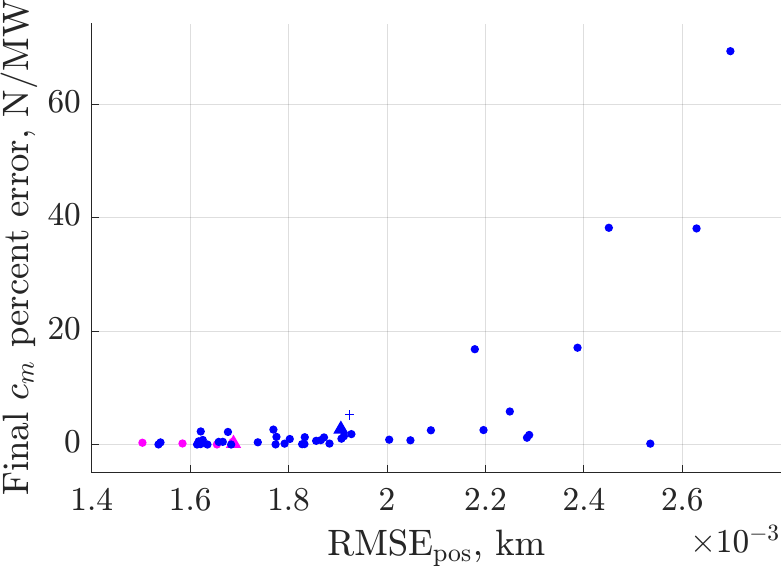}
        \caption{View of final $c_\text{m}$ percent error versus position RMSE clustering.}
        \label{fig:num2_metricClust_xz}
    \end{subfigure}
    \caption{Performance metric space clustering of all out-of-plane L2D engagement cases.}
    \label{fig:num2_metricClust_ALL}
\end{figure}

Results of the $k$-means clustering in the performance metric space are shown in Fig.~\ref{fig:num2_metricClust_ALL}. Cluster 2 comprises cases with total periapsis decrease magnitudes greater than all cases within cluster 1, where each of cluster 2's cases also exhibits relatively low magnitude position RMSE values. Additionally, all cases within cluster 2 provide exceptionally low final $c_\text{m}$ percent errors by the end of their respective L2D engagements. Cluster 1 comprises the majority of the total cases analyzed in this experiment, where several outlier cases exhibit large magnitude position RMSE values and final $c_\text{m}$ percent errors. These outliers were also cases with much shorter L2D engagement intervals that resulted in an order of magnitude less total periapsis radius decrease magnitude as well. The following general trends can be extracted from the clustering results: (1) position RMSE magnitudes decrease as the total periapsis radius decrease magnitude increases; (2) the final $c_\text{m}$ percent error decreases as the total periapsis radius decrease magnitude increases; (3) the final $c_\text{m}$ percent error increases as the position RMSE magnitude increases. These trends are expected following the coplanar analysis, where cases with a larger periapsis decrease magnitude generally resulted in a relatively low magnitude position RMSE by the end of an L2D engagement; however, the out-of-plane clustering shows that well-performing $c_\text{m}$ estimation is not guaranteed.

\begin{figure}[!h]
    \centering
    \begin{subfigure}{.49\linewidth}
        \centering
        \includegraphics[width=\linewidth]{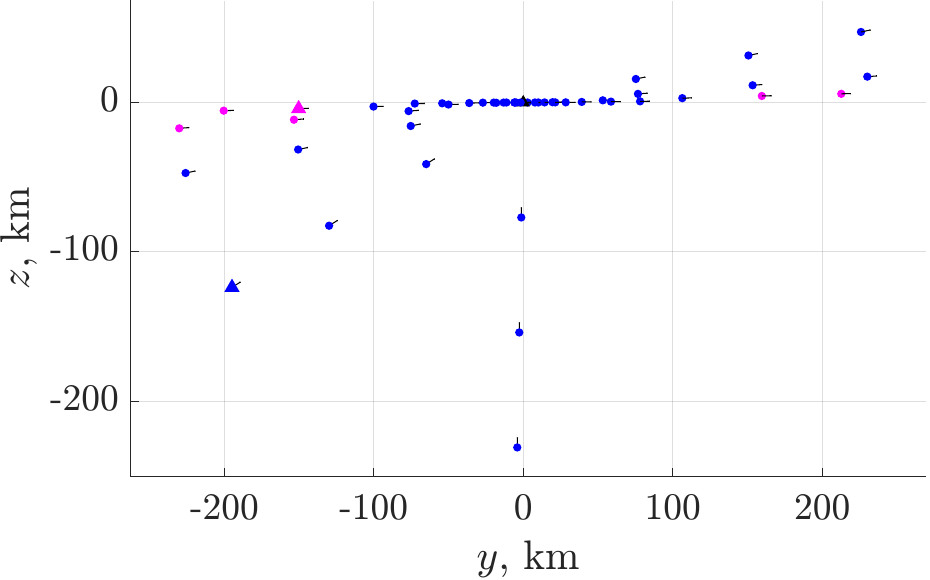}
        \caption{$yz$-plane view.}
        \label{fig:num2_posClust_yz}
    \end{subfigure}
    \begin{subfigure}{.49\linewidth}
        \centering
        \includegraphics[width=\linewidth]{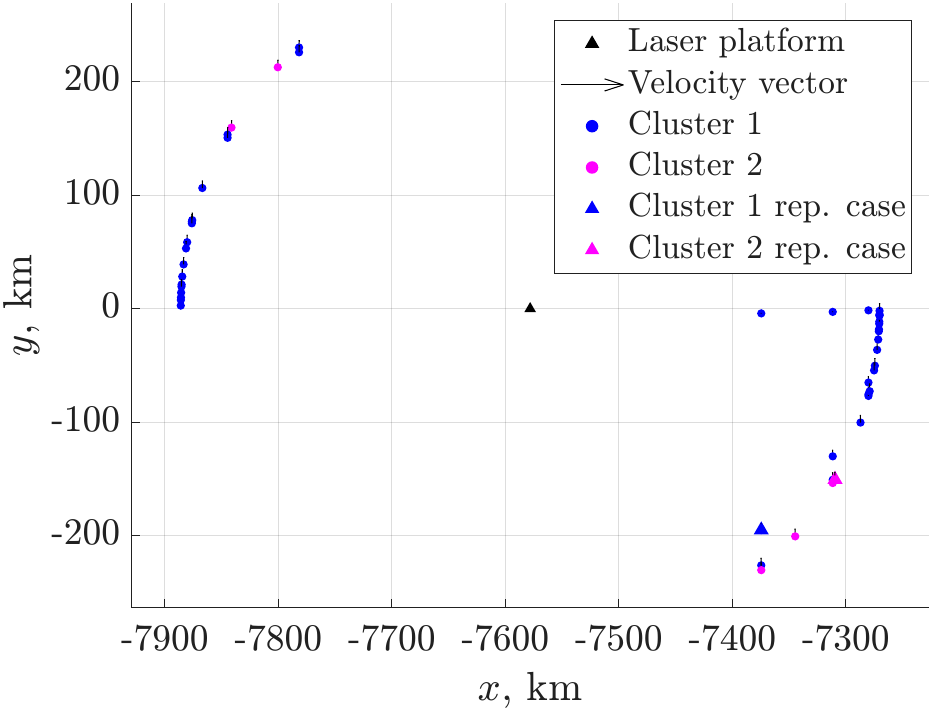}
        \caption{$xy$-plane view.}
        \label{fig:num2_posClust_yx}
    \end{subfigure}
    \caption{Position space clustering of all out-of-plane cases given in the ECI coordinate frame at the first time step where an L2D engagement could begin.}
    \label{fig:num2_posClust_ALL}
\end{figure}

With the general trends of the metric space in mind, the two clusters are translated to the more familiar position space as shown in Fig.~\ref{fig:num2_posClust_ALL}. The position space clustering elucidates how specific encounter geometries in out-of-plane L2D engagements lead to overall improvements across all performance metrics. In this representation, all cases within cluster 2 are seen to be close to coplanar with the laser platform, while also being initialized with a smaller magnitude relative altitude with respect to the laser platform. The coplanar analysis showed that all reducing the relative altitude improved the overall performance of an L2D engagement; however, the relative inclination between the laser platform and debris appears to play a critical role in the overall performance of the out-of-plane L2D engagements. To better understand the impacts of these different encounter geometries, an in-depth analysis is conducted with respect to a representative case from both clusters to identify driving factors that contribute to improving or degrading the overall performance of an out-of-plane L2D engagement.

\subsubsection{Target Debris Periapsis Radius Decrease and Linear System Observability Analysis} \label{subsubsec:exp_2_peri_obs}

\begin{figure}[!h]
    \centering
    \begin{subfigure}{\linewidth}
        \centering
        \includegraphics[width=0.8\linewidth]{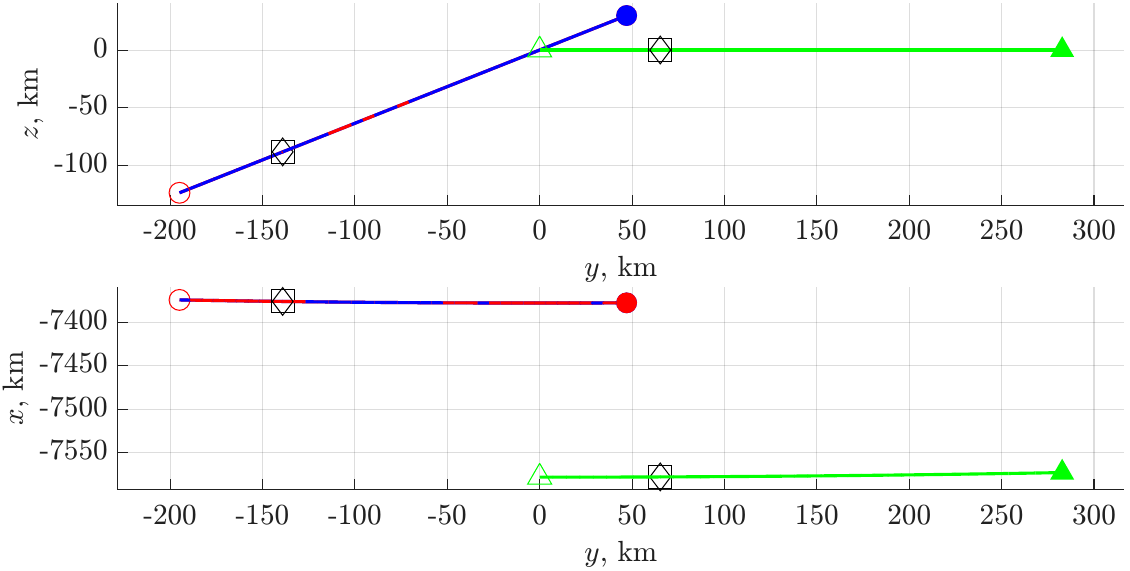}
        \caption{Cluster 1 representative case.}
        \label{fig:num2_traj_out_poor}
        \end{subfigure}
        \begin{subfigure}{\linewidth}
        \centering
        \includegraphics[width=0.8\linewidth]{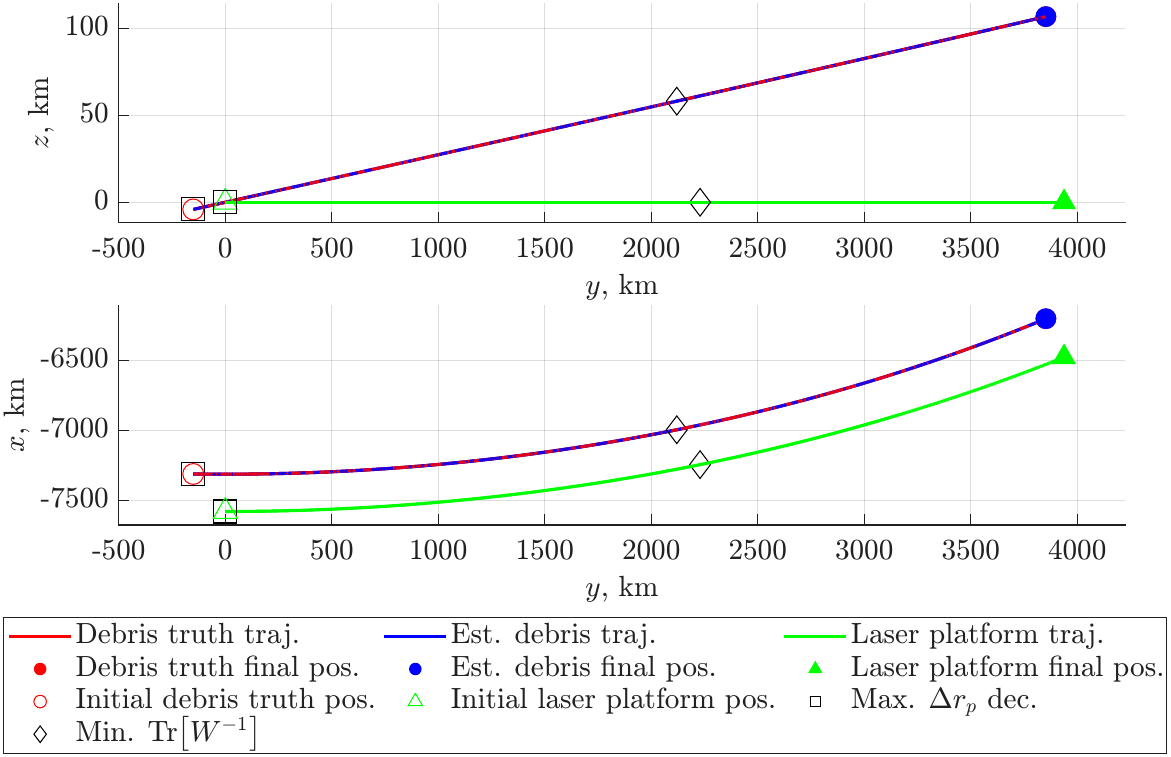}
        \caption{Cluster 2 representative case.}
        \label{fig:num2_traj_out_well}
    \end{subfigure}
    \caption{Visualization of the out-of-plane cluster 1 and 2 representative cases' target debris and laser platform propagated trajectories during the L2D engagement.}
\end{figure}

\begin{figure}[!h]
    \centering
    \begin{subfigure}{0.49\linewidth}
        \centering
        \includegraphics[width=\linewidth]{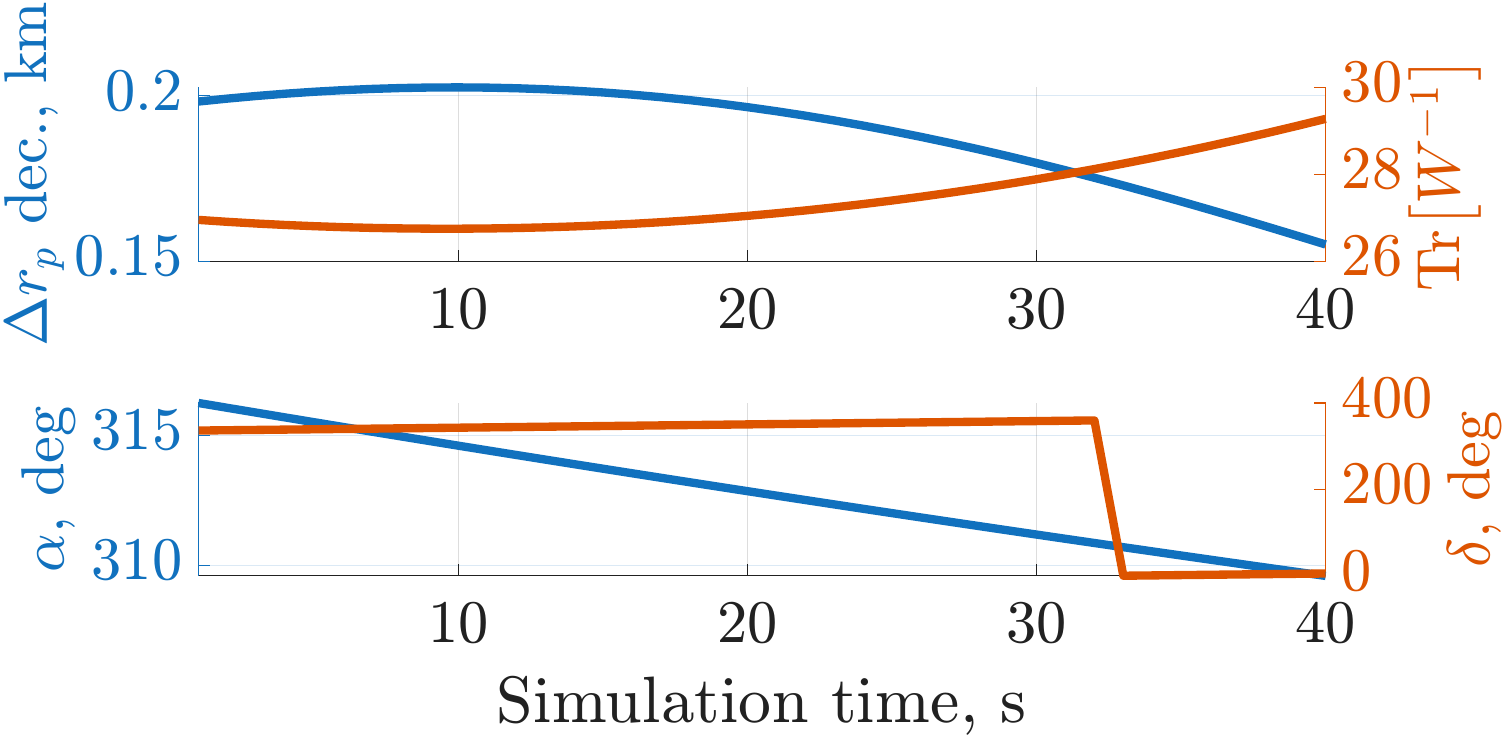}
        \caption{Cluster 1 representative case.}
        \label{fig:num2_peri_obs_poor}
    \end{subfigure}
    \hfill
    \begin{subfigure}{0.49\linewidth}
        \centering
        \includegraphics[width=\linewidth]{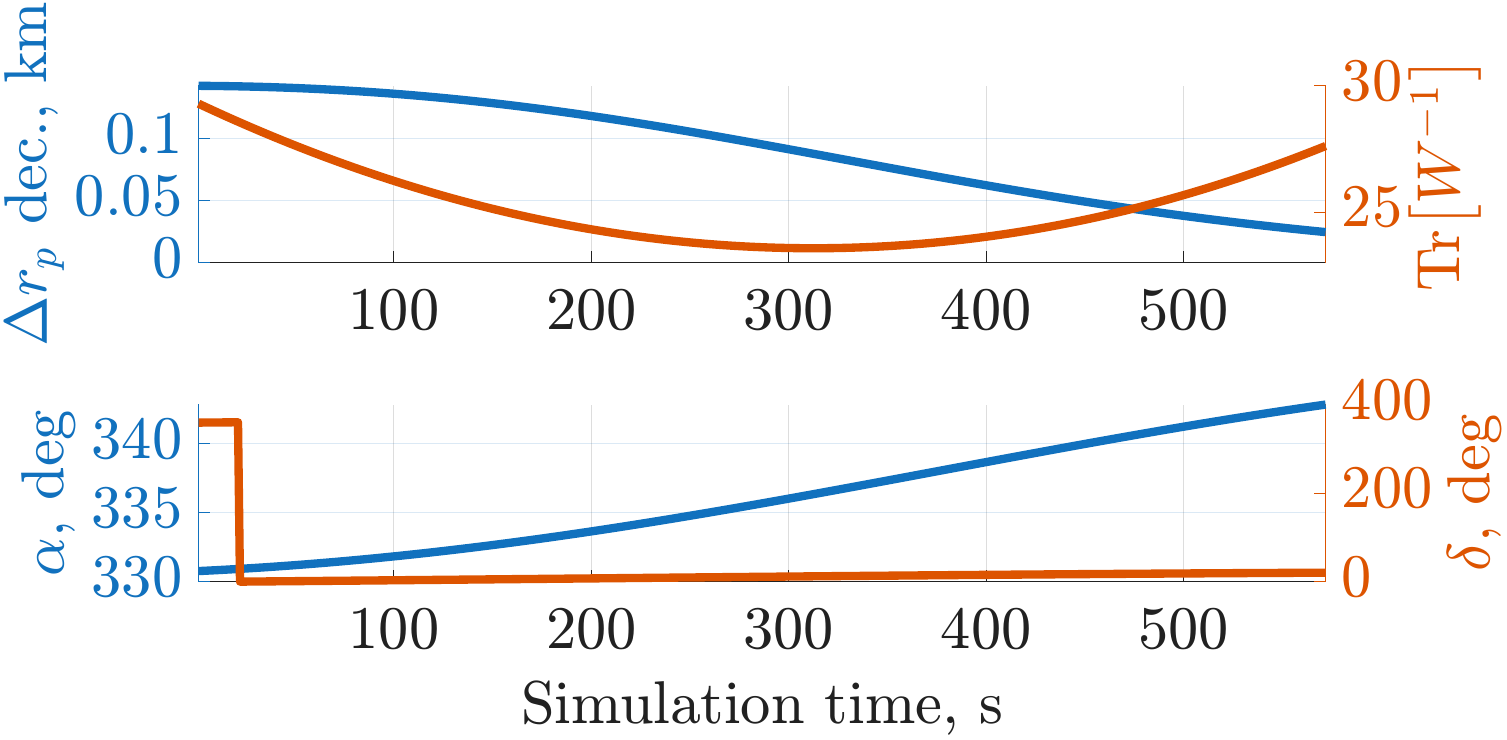}
        \caption{Cluster 2 representative case.}
        \label{fig:num2_peri_obs_well}
    \end{subfigure}
    \caption{Periapsis radius decrease and observability Gramian inverse matrix trace variation over the simulation time length for the cluster 1 and 2 representative cases.}
\end{figure}

Cluster 1 and 2 representative cases' target debris and laser platform orbits and propagated states are visualized in Figs.~\ref{fig:num2_traj_out_poor} and \ref{fig:num2_traj_out_well}, respectively. The cluster 1 and 2 representative cases are propagated forward in time with a \SI{1}{s} time step size for a total simulation time length of \SI{39.0}{s} and \SI{571.0}{s}, respectively. Changes in the periapsis radius decrease magnitude over time for the cluster 1 and 2 representative cases' target debris are shown in Fig.~\ref{fig:num2_peri_obs_poor} and Fig.~\ref{fig:num2_peri_obs_well}, respectively. The cluster 1 and 2 representative cases reduced the periapsis radius of their respective target debris by \SI{7.5588}{km} and \SI{52.1400}{km}, respectively. As compared to the coplanar cases, the out-of-plane L2D engagements are much shorter; thus, the total periapsis radius decrease magnitude at the end of a single out-of-plane L2D engagement is generally less than that of the analyzed coplanar cases. The $\text{Tr}\left[W^{-1}\right]$ for the cluster 2 representative case has a lower magnitude throughout the majority of the L2D engagement compared to the cluster 1 representative case, which is due to a greater change in angle measurements over the full simulation time.

\subsubsection{EKF Performance} \label{subsubsec:exp_2_ekf}
\begin{figure}[!h]
    \centering
    \begin{subfigure}{0.49\linewidth}
        \centering
        \includegraphics[width=\linewidth]{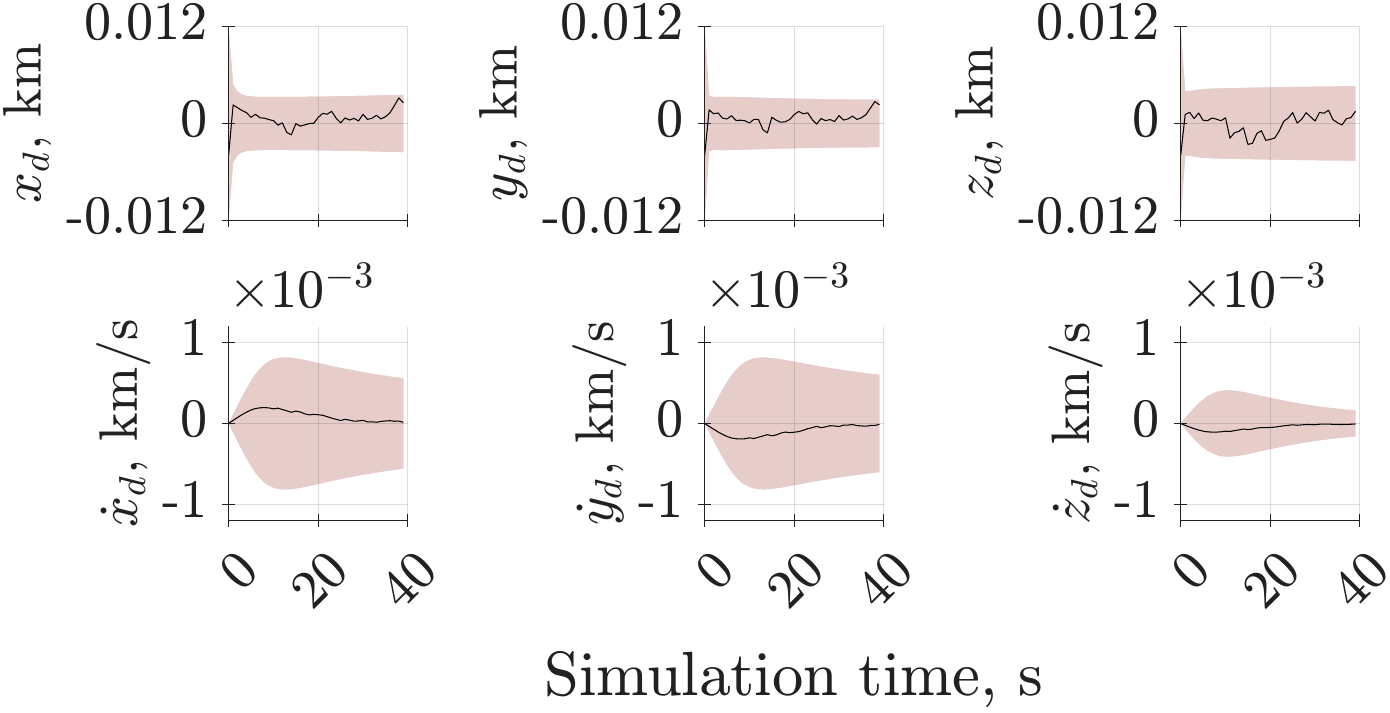}
        \caption{Cluster 1 representative case.}
        \label{fig:num2_state_err_poor}
    \end{subfigure}
    \hfill
    \begin{subfigure}{0.49\linewidth}
        \centering
        \includegraphics[width=\linewidth]{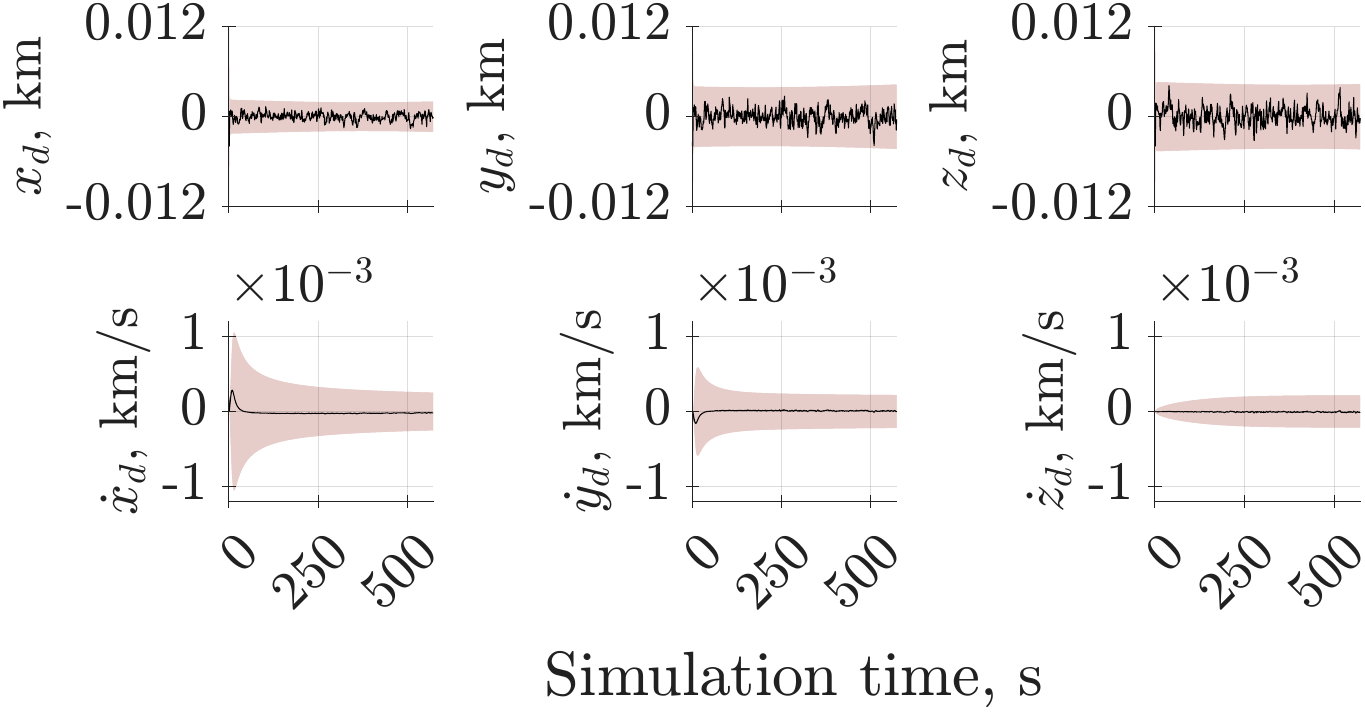}
        \caption{Cluster 2 representative case.}
        \label{fig:num2_state_err_well}
    \end{subfigure}
    \caption{Cluster 1 and 2 representative cases' state errors (black) and corresponding three standard deviation bounds (red) given over their respective simulation time length.}
\end{figure}

The state error and corresponding three standard deviation bounds over the simulation time length for the cluster 1 and 2 representative cases are given in Fig.~\ref{fig:num2_state_err_poor} and Fig.~\ref{fig:num2_state_err_well}, respectively. Considering all measurement time steps until the end of their respective L2D engagements, the position RMSE of the cluster 1 and 2 representative cases are \SI{1.9061}{m} and \SI{1.6880}{m}, respectively. The state estimation performance between the two cases is similar, where the accurate estimation of the $x$-position and extended simulation time length in the cluster 2 representative case contribute to a lower-magnitude position RMSE by the end of the L2D engagement as compared to the cluster 1 representative case.

\begin{figure}[!h]
    \centering
    \begin{subfigure}{0.49\linewidth}
        \centering
        \includegraphics[width=\linewidth]{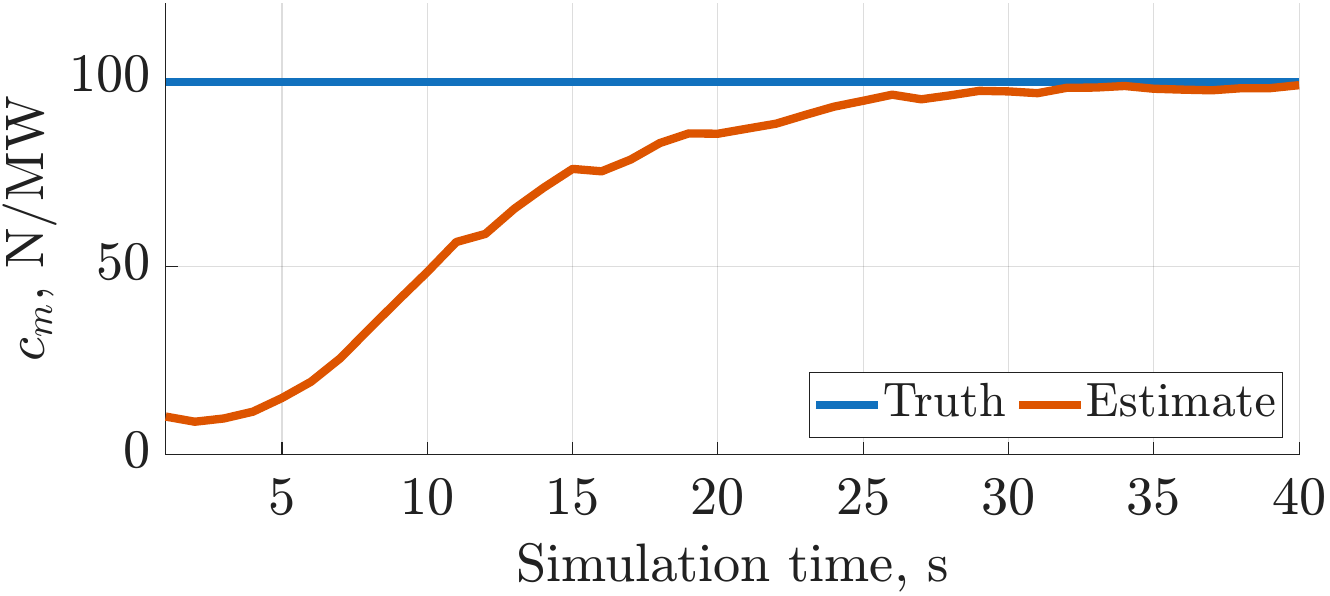}
        \caption{Cluster 1 representative case.}
        \label{fig:num2_para_poor}
    \end{subfigure}
    \hfill
    \begin{subfigure}{0.49\linewidth}
        \centering
        \includegraphics[width=\linewidth]{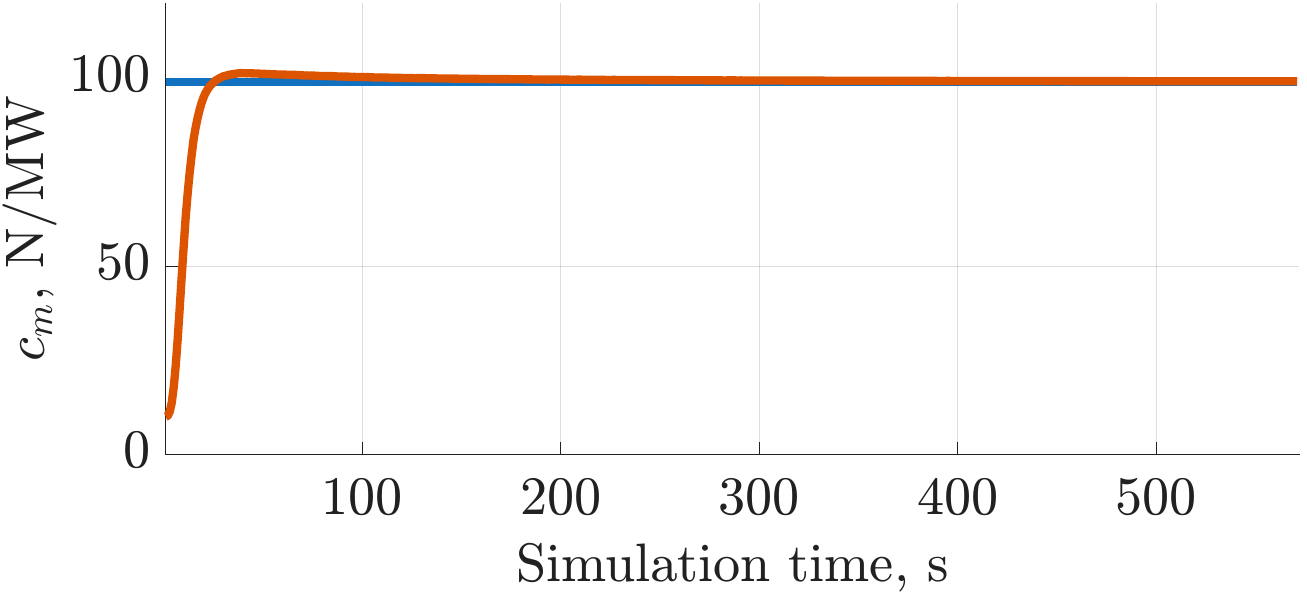}
        \caption{Cluster 2 representative case.}
        \label{fig:num2_para_well}
    \end{subfigure}
    \caption{$c_\text{m}$ estimate and $c_\text{m}$ truth over the cluster 1 and 2 representative cases' simulation time lengths.}
    \label{fig:num2_para_ALL}
\end{figure}

Both of the representative cases approach the true $c_\text{m}$ value by the end of the L2D engagement, as shown in Fig.~\ref{fig:num2_para_ALL}, where the final $c_\text{m}$ percent difference for the first and second representative case's are \SI{2.6992}{\%} and \SI{0.2014}{\%}, respectively. While cluster 1's representative case has a much shorter L2D engagement, the $c_\text{m}$ estimate is still able to be relatively close to the true $c_\text{m}$, which is attributed to the larger-magnitude change in the debris's periapsis radius over a short time span. This result highlights that $c_\text{m}$ estimation performance is greatly impacted by the magnitude of the debris's perturbation due to laser ablation during an L2D engagement.

\section{Numerical Experiment 3: Parametric Laser System Analysis} \label{sec:num_exp_3_param}
This numerical experiment investigates how observability of linear systems, nonlinear state and parameter estimation, and deorbiting performance are affected by changes in laser system parameters. The laser systems considered include the International Coherent Amplification Network~(ICAN) \cite{SOULARD2014192} and Laser Ablative Debris Removal by Orbital Impulse Transfer (L'ADROIT)~\cite{phipps2014Ladroit} conceptual laser platforms. Additionally, commercial laser systems capable of laser ablation may be adapted for orbital debris remediation in space-based laser ablation \cite{nanofluxMM} and are therefore considered in this numerical experiment.

\subsection{Experimental Setup} \label{subsec:exp_set_3_param}

This analysis assumes three sets of laser parameters varying from low to high pulse energy of the proposed laser while in a shooting mode. The target debris for each case is assumed to be a non-rotating homogeneous solid aluminum sphere with a characteristic length of \SI{5}{cm}. The low-magnitude pulse-energy system is selected from a commercial off-the-shelf (COTS) laser system, while the medium-and high-pulse-energy laser systems are given as the ICAN and L'ADROIT laser systems, respectively. The set of laser parameters for each assumed laser system is provided in Table~\ref{tab:low_med_high_para}.
\begin{table}[htbp]
    \caption{Low-, medium-, and high-pulse energy laser system parameters.}
    \centering
\begin{threeparttable}
    \begin{tabular}{lrrrr}
    \hline
    \hline
    Parameter & Low $E$: COTS & Medium $E$: ICAN & High $E$: L'ADROIT & Unit \\
    \hline
    Pulse energy, $E$ & 5.0 & 100.0 & 380.0 & \si{J}\\
    Wavelength, $\lambda$ & 532.0 & 1000.0 & 355.0 & \si{nm}\\
    Pulse duration, $\tau$ & 5.0 & 0.10 & 0.10 & \si{ns}\\
    Pulse repetition frequency, $\upsilon$ & 10.0 & 1000 & 56.0 & \si{Hz}\\
    Primary mirror diameter, $D$ & 3.0 & 3.0 & 1.5 & \si{m}\\
    Beam Quality, $M^2$ & 5.0 & 1.0 & 2.0 & -\\
    Optimal laser intensity & \num{1.202e3} & 8500 & 8500 & \si{MW/cm^2}\\
    Optimal delivered fluence, $\Psi_{\text{opt}}$ & 60.10 & 8.50 & 8.50 & \si{kJ/m^2}\\
    Optimal $c_\text{m}$\tnote{1} & 99.00 & 99.00 & 99.00 & \si{N/MW}\\
    Time-averaged intensity, $\bar{I}$ & 601.0 & 8500.0 & 476.0 & \si{kW/m^2}\\
    Plasma ignition fluence, $\Psi_{\text{p}}$ & 31.82 & 4.50 & 4.50 & \si{kJ/m^2}\\
    Max. ablation distance, $L_{\text{max}}$\tnote{1} & 5.83 & 217.08 & 307.78 & \si{km}\\
    Min. ablation distance, $L_{\text{min}}$\tnote{1} & 4.65 & 65.12 & 92.33 & \si{km} \\
    Beam spatial profile & Super-Gaussian & Gaussian & Gaussian & -\\
    Diffraction constant, $c_\text{diff}$ & 1.5 & $\frac{4}{\pi}$ & $\frac{4}{\pi}$ & - \\
    Output burst power & 0.05 & 100.0 & 21.0 & \si{kW}\\
    Overall laser e-o efficiency, $T_{\text{eff}}$ & 0.30 & 0.30 & 0.32 & -\\
    Input burst power & 0.17 & 333.33 & 70.0 & \si{kW}\\
    \hline
    \hline
    \end{tabular}
    \begin{tablenotes}
    \item[1] \footnotesize{These noted variables are calculated according to the equations presented in this section and differ from the presented results in the ICAN and L'ADROIT referenced work.}\\
    \end{tablenotes}
\end{threeparttable}
    \label{tab:low_med_high_para}
\end{table}

The COTS laser system is not provided for explicit use as a space-based laser that will perform orbital debris remediation, so additional laser and laser-matter interaction parameters must be defined to allow for an analysis of the system's performance. The maximum distance that laser ablation can occur, $L_{\text{max}}$, is determined using Eq.~\eqref{eq:max_range}, which is derived from the integration of pertinent equations presented in Refs.~\cite{Stafe_LA_book_2014,phipps_bonnal_2016,phipps2014Ladroit}, 

\begin{equation}
    L_{\text{max}} = \sqrt{\frac{E T_\text{eff}}{\pi \Psi_{\text{p}}}} \frac{2D}{c_\text{diff} M^2 \lambda} \label{eq:max_range}
\end{equation}
where the laser system's pulse energy and wavelength are denoted as $E$ and $\lambda$, respectively. The overall electro-optical efficiency is denoted as $T_\text{eff}$, and the primary mirror diameter is denoted as $D$. The laser beam quality and diffraction constant are denoted as $M^2$ and $c_\text{diff}$, respectively. The minimum fluence needed to induce plasma ignition is denoted as $\Psi_\text{p}$.

According to experimental data and models predicting the formation of plasma above the ablated surface of metals in a vacuum, the minimum laser intensity required to induce plasma formation due to laser ablation is assumed to adhere to the relationship given in Eq.~\eqref{eq:intensity_min_plasma}.

\begin{equation}
    I_p \sqrt{\tau} \approx B_p \label{eq:intensity_min_plasma}
\end{equation}
In this relationship, the laser intensity needed for plasma ignition is denoted as $I_\text{p}$, and $B_\text{p}$ is a material-dependent coefficient relating $I_\text{p}$ to the pulse duration denoted as $\tau$. For an aluminum target, the material-dependent coefficient is assumed to be approximately equal to \num{4.5e4}~\unit{W~s^{1/2}/cm^2} \cite{Stafe_LA_book_2014}.
The minimum fluence needed to induce plasma formation is then calculated using Eq.~\eqref{eq:fluence_min_plasma},

\begin{equation}
    \Psi_{\text{p}} \approx B_{\text{p}}\sqrt{\tau} \label{eq:fluence_min_plasma}
\end{equation}
where the optimal delivered fluence, $\Psi_{\text{opt}}$ is calculated considering the relationship shown in Eq.~\eqref{eq:fluence_opt},

\begin{equation}
    \Psi_{\text{opt}} \approx B\sqrt{\tau} \label{eq:fluence_opt}
\end{equation}
where $B$ is a material-dependent coefficient equal to \num{8.5e8}~\si{W~s^{1/2}/m^2} \cite{phipps2014Ladroit}.

A single target debris and laser platform pair is selected for this analysis, where the target debris initialization is chosen such that an L2D engagement will be possible for all laser systems considered in this numerical experiment. Using the same target debris orbit, each case's target debris is initialized such that the next time step would be the first possible time step for an L2D engagement, considering each laser system's unique minimum engagement distance. This minimum engagement distance is defined as the engagement distance resulting in a laser fluence equal to \num{50}~\si{kJ/m^2}, which has been shown in numerical experiments to produce splashing of large ejecta particles during laser ablation of aluminum \cite{Stafe_LA_book_2014}. The initial position and velocity for the laser platform and each target debris used in this numerical experiment are given in Table~\ref{tab:exp3_KOE}.
\begin{table}[!h]
    \caption{Target debris and laser platform initial position and velocities used for the parametric analysis.}
    \centering
    \begin{tabular}{lrrrrrr}
        \hline
        \hline
            Initialized object & $x$, \si{km} & $y$, \si{km} & $z$, \si{km} & $\dot{x}$, \si{km/s} & $\dot{y}$, \si{km/s} & $\dot{z}$, \si{km/s}  \\
            \hline 
            COTS target debris & -7574.3 & -4.3687 & -0.0434 & -0.0042 & 7.2540 & 0.0721  \\
            ICAN target debris & -7571.2 & -216.9581 & -2.1555 & -0.2078 & 7.2510 & 0.0720  \\
            L'ADROIT target debris & -7568.0 & -307.5976 & -3.0560 & -0.2946 & 7.2480 & 0.0720  \\
            Laser platform & -7578.1 & 0.0 & 0.0 & 0.0 & 7.2525 & 0  \\
        \hline
        \hline
    \end{tabular}
    \label{tab:exp3_KOE}
\end{table}

\subsection{Experiment Results} \label{subsec:results_3_param}
\begin{figure}
    \centering
    \includegraphics[width=0.8\linewidth]{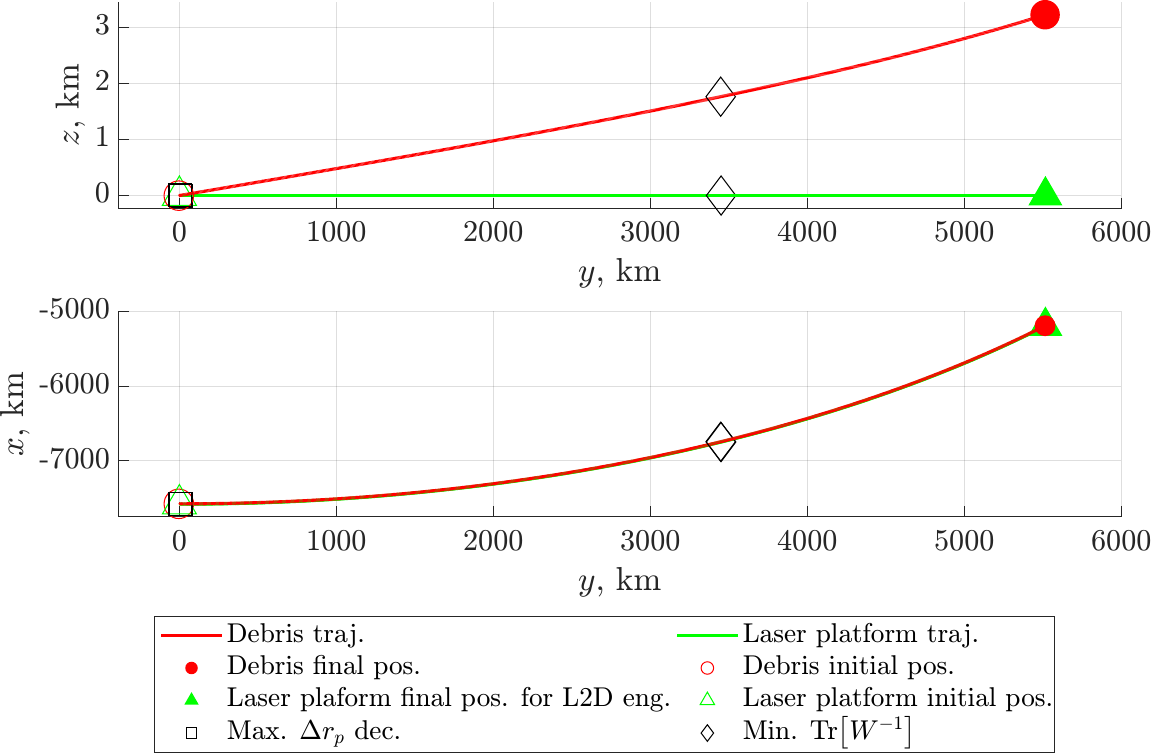}
    \caption{Visualization of the target debris and laser platform orbits and a detailed view of propagated states during the COTS laser system case's L2D engagement.}
    \label{fig:cots_isometric}
\end{figure}

\begin{figure}
    \centering
    \includegraphics[width=0.8\linewidth]{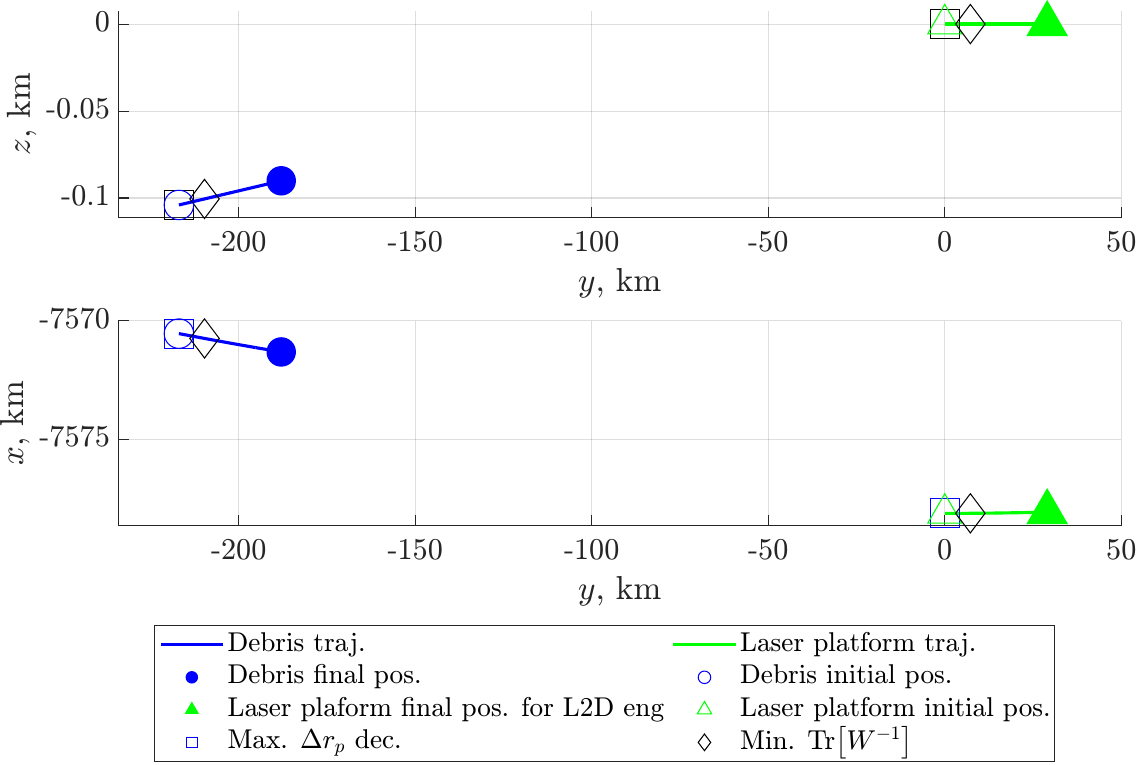}
    \caption{Visualization of the target debris and laser platform orbits and a detailed view of propagated states during the ICAN laser system case's L2D engagement.}
    \label{fig:ican_isometric}
\end{figure}

\begin{figure}
    \centering
    \includegraphics[width=0.8\linewidth]{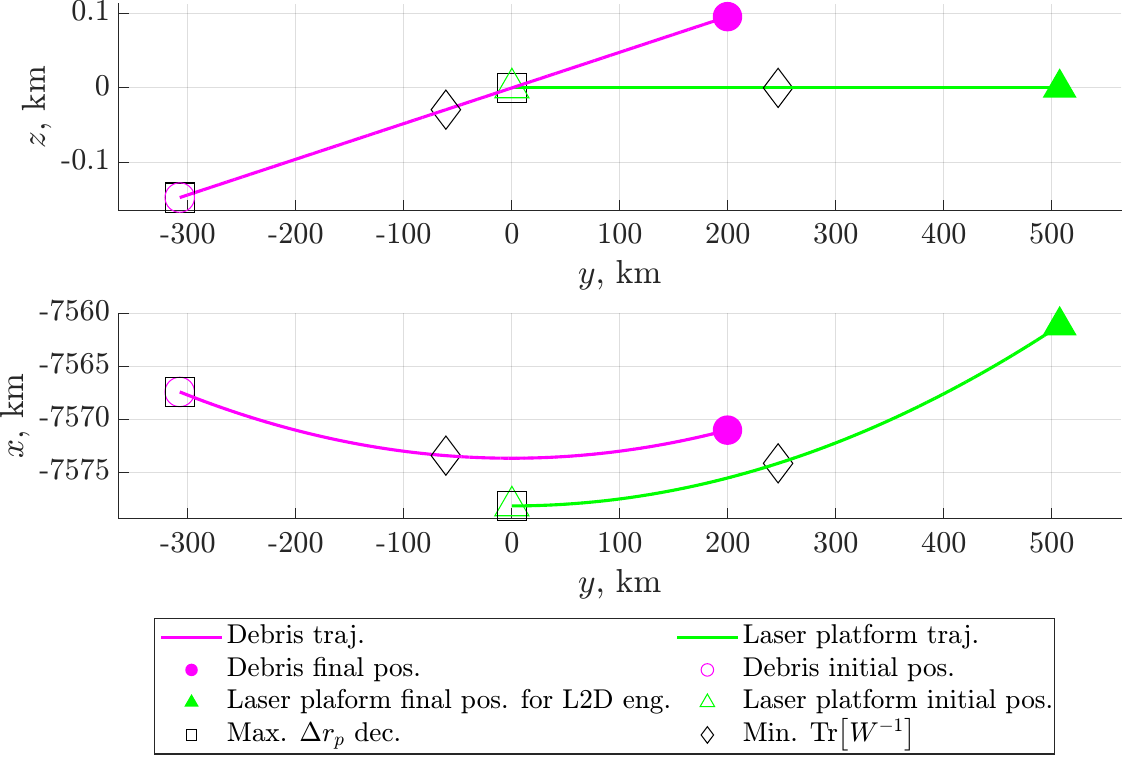}
    \caption{Visualization of the target debris and laser platform orbits and a detailed view of propagated states during the L'ADROIT laser system case's L2D engagement.}
    \label{fig:ladroit_isometric}
\end{figure}

A comparison of three laser systems considered in this experiment is presented for the target debris periapsis radius decrease, linear system observability, and nonlinear state and parameter estimation performance. Figures~\ref{fig:cots_isometric}, \ref{fig:ican_isometric}, and \ref{fig:ladroit_isometric} provide a visualization of the laser platform, the target debris orbit, and propagated states during each case's respective L2D engagements, respectively.

\subsubsection{Target Debris Periapsis Radius Decrease and Linear System Observability} \label{subsubsec:exp_3_peri_obs}
\begin{figure}[!h]
    \centering
    \begin{subfigure}{0.7\linewidth}
        \centering
        \includegraphics[width=\linewidth]{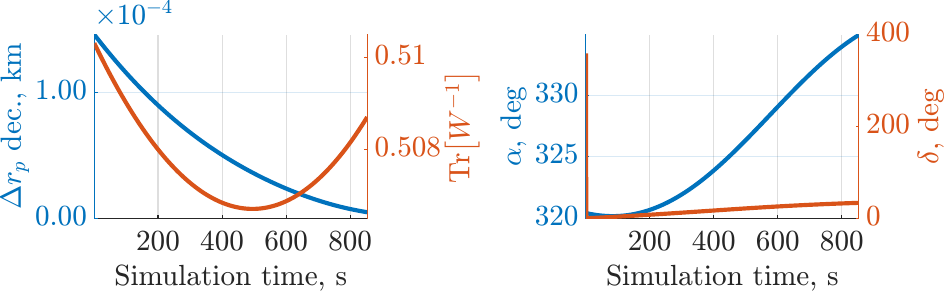}
        \caption{COTS laser system.}
        \label{fig:num3_COTS_peri_obs}
    \end{subfigure}
    \hfill
    \begin{subfigure}{0.7\linewidth}
        \centering
        \includegraphics[width=\linewidth]{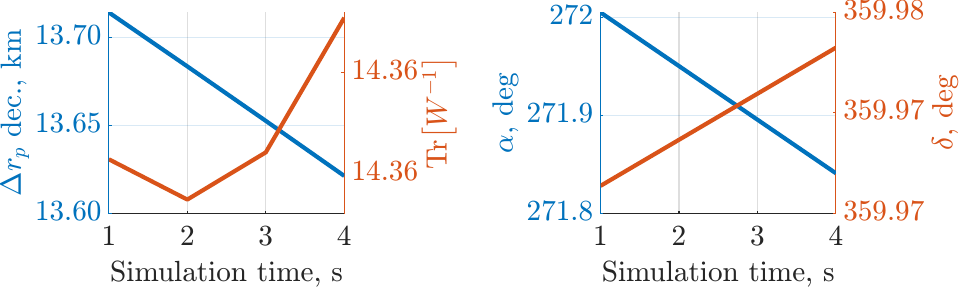}
        \caption{ICAN laser system.}
        \label{fig:num3_ICAN_peri_obs}
    \end{subfigure}
     \hfill
    \begin{subfigure}{0.7\linewidth}
        \centering
        \includegraphics[width=\linewidth]{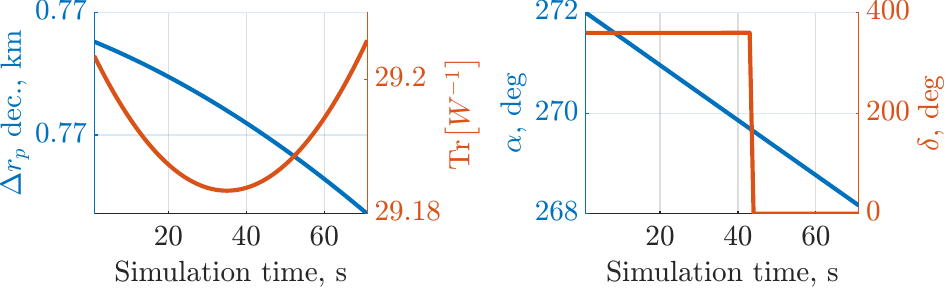}
        \caption{L'ADROIT laser system.}
        \label{fig:num3_LADROIT_peri_obs}
    \end{subfigure}
    \caption{Periapsis radius decrease and observability Gramian inverse matrix trace variation over the simulation time length for the COTS, ICAN, and L'ADROIT laser system cases.}
\end{figure}

The periapsis radius decrease and $\text{Tr}\left[W^{-1}\right]$ magnitudes at each time step for the three laser system cases are shown in Figs.~\ref{fig:num3_COTS_peri_obs}, \ref{fig:num3_ICAN_peri_obs}, and \ref{fig:num3_LADROIT_peri_obs}, respectively. The total target debris periapsis radius reduction by the end of the simulation time for the COTS, ICAN, and L'ADROIT cases is \SI{0.0475}{km}, \SI{54.6715}{km}, and \SI{54.3642}{km}, respectively. The ICAN system has a $\Delta r_p$ decrease magnitude several orders of magnitude larger than the COTS and L'ADROIT laser systems due to the ICAN system having a much larger time-averaged laser intensity as compared to the other laser systems; consequently, the ICAN system is able to decrease the target debris periapsis radius by a similar value to the L'ADROIT system in less than one-tenth of the time over a single L2D engagement. The COTS laser system is seen to have the lowest magnitude $\text{Tr}\left[W^{-1}\right]$, and the L'ADROIT laser system has the largest magnitude $\text{Tr}\left[W^{-1}\right]$ across all cases. Additionally, the COTS laser system has a $\text{Tr}\left[W^{-1}\right]$ magnitude that is two orders of magnitude lower than the ICAN and L'ADROIT laser system cases. This lower overall $\text{Tr}\left[W^{-1}\right]$ for the COTS system is attributed to the significantly larger change in the elevation angle over the simulation time length; however, the ICAN system also has a $\text{Tr}\left[W^{-1}\right]$ magnitude that is nearly half of the L'ADROIT $\text{Tr}\left[W^{-1}\right]$ magnitude while the variation in the measurement angles is relatively small. This reduction in $\text{Tr}\left[W^{-1}\right]$ emphasizes how a sufficiently large-magnitude change in the periapsis can improve overall observability of the system's state variables despite having a small variation in the measurement angles.

\subsubsection{EKF Performance} \label{subsubsec:exp_3_ekf}

\begin{figure}[!h]
    \centering
    \begin{subfigure}{0.49\linewidth}
        \centering
        \includegraphics[width=\linewidth]{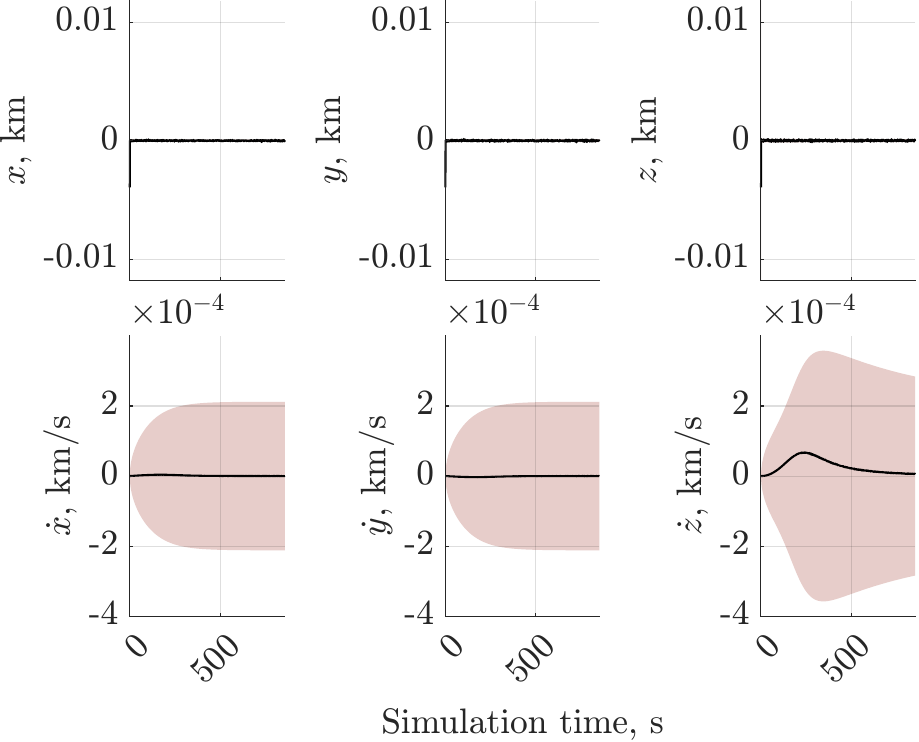}
        \caption{COTS laser system.}
        \label{fig:num3_COTS_state_err}
    \end{subfigure}
    \hfill
    \begin{subfigure}{0.49\linewidth}
        \centering
        \includegraphics[width=\linewidth]{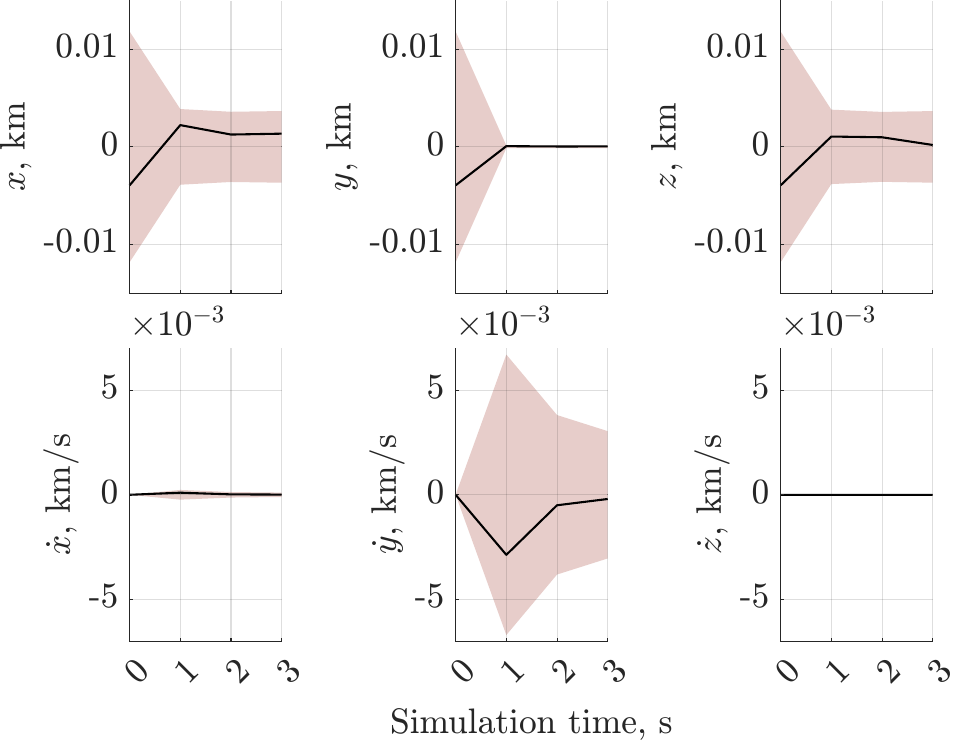}
        \caption{ICAN laser system.}
        \label{fig:num3_ICAN_state_err}
    \end{subfigure}
     \hfill
    \begin{subfigure}{0.49\linewidth}
        \centering
        \includegraphics[width=\linewidth]{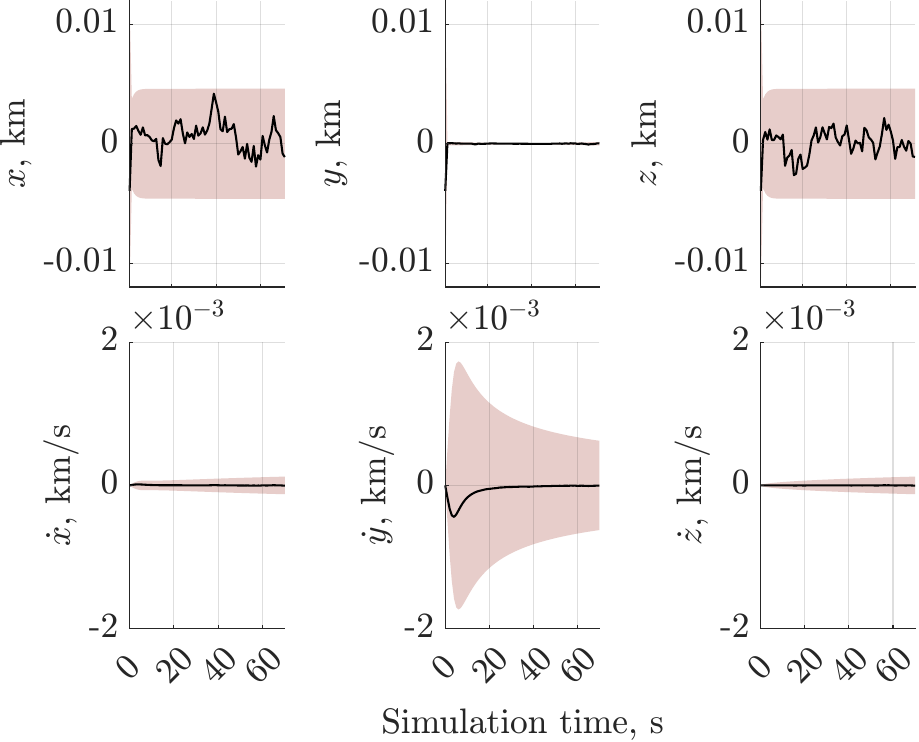}
        \caption{L'ADROIT laser system.}
        \label{fig:num3_LADROIT_state_err}
    \end{subfigure}
    \caption{Target debris state errors (black) and corresponding three standard deviation bounds (red) given over the simulation time length.}
\end{figure}

\begin{figure}[!h]
    \centering
    \includegraphics[width=.7\linewidth]{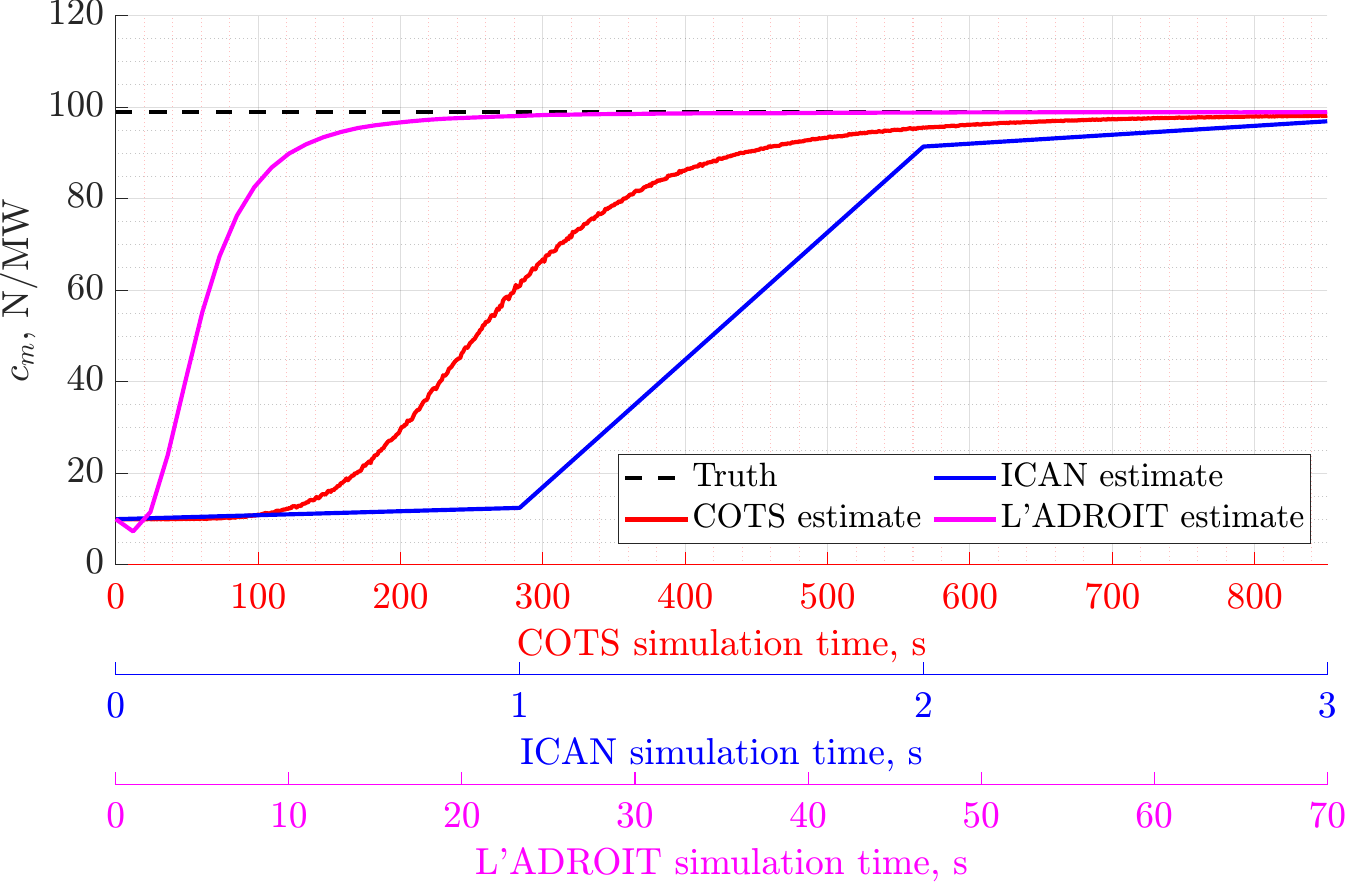}
    \caption{Comparison of the parameter estimation performance for each laser system considered in the parametric analysis.}
    \label{fig:num3_para_err_comp}
\end{figure}

The state errors and corresponding three standard deviation bounds for each laser system plotted over the simulation time length for the COTS, ICAN, and L'ADROIT laser system cases are shown in Figs.~\ref{fig:num3_COTS_state_err}, \ref{fig:num3_ICAN_state_err}, and \ref{fig:num3_LADROIT_state_err}, respectively. The position RMSE for the COTS, ICAN, and L'ADROIT at the end of the simulation time length are \num{2.4309e-4} \si{km}, \SI{0.0038}{km}, and \SI{0.0019}{km}, respectively. The COTS system is seen to have significantly less state error and tighter $\pm 3\sigma$ bounds on position states compared to the ICAN and L'ADROIT systems, but larger errors and broader $\pm 3\sigma$ bounds are observed for the COTS velocity states. The parameter estimation for each laser system case is shown in Fig.~\ref{fig:num3_para_err_comp}, where the $c_\text{m}$ estimation given by the COTS case is shown to have a slow rise to the truth value over the simulation time length. Figure~\ref{fig:num3_para_err_comp} also shows the ability for the ICAN system to quickly converge to the $c_\text{m}$ truth, despite having a significantly shorter L2D engagement interval compared to the other two cases.

The results from this numerical experiment show that a smaller power laser system has the potential to provide more accurate state estimates while trading off the capacity to quickly estimate $c_\text{m}$ and lower the target debris periapsis radius during a single L2D engagement. Therefore, smaller laser systems may be better suited for just-in-time collision avoidance, where a slight perturbation in the target debris states would decrease the probability of collision with high-value assets by slightly altering the debris's orbit while also leading to increased accuracy of target debris state estimates. On the other hand, a larger magnitude perturbation from a larger laser system is shown to have a position RMSE within \SI{4}{m}, which suggests that a single L2D engagement also has the potential to significantly reduce the target debris periapsis radius if a slightly larger state error margin is allowable during a remediation campaign.

\section{Recommendations and Limitations} \label{sec:rec_lim}

The authors recommend considering the following key insights based on observations drawn from each numerical experiment:

\begin{enumerate}
    \item In general, the total debris periapsis decrease magnitude can be increased if the relative inclination, relative altitude, or both the relative inclination and altitude of the debris with respect to the laser platform are decreased. Reducing either of these relative characteristics results in an encounter geometry that provides a longer L2D engagement, while also increasing the debris's periapsis reduction magnitude at each time step of the engagement due to the L2D perturbing acceleration vector being closer to anti-parallel with the debris's velocity vector. Significantly altering a laser platform's orbital inclination via plane change maneuvers may require an associated significant $\Delta V$ magnitude, so this recommendation is given with respect to initial laser platform deployment or maneuvering to reasonably nearby orbital inclinations available during the operational lifetime of the laser platform.
    \item The estimation performance increases as the debris's periapsis radius decrease magnitude and angle measurement rate of change increase, regardless of the given laser system and across all coplanar and out-of-plane encounter geometries. In general, the rate of change of the angle measurements dominates the increase in estimation performance; however, as observed through the ICAN laser system analysis, a sufficiently large periapsis radius decrease magnitude can provide low-magnitude $\text{Tr}\left[W^{-1}\right]$ values during L2D engagements with small variations in the angle measurements. While the performance of the ICAN system in Numerical Experiment 3 suggests that the debris states and $c_\text{m}$ can still be estimated relatively accurately during a short L2D engagement, the authors recommend that future practitioners take caution in the implementation of such a high-power laser system to ensure that the laser platform does not lose custody of the debris when it is significantly perturbed over a short period of time.
    \item When prioritizing $c_\text{m}$ estimate convergence, it is recommended to utilize a laser system with a higher magnitude time-averaged laser intensity. As shown in Sec.~\ref{sec:num_exp_3_param}, all L2D engagements with various laser system parameters are capable of providing $c_\text{m}$ estimates that converge towards the true $c_\text{m}$ value, but the time to approach the true value increased as the laser system's time-averaged laser intensity decreased; on the other hand, the low magnitude time-averaged laser intensity still provides a slight change in the debris's orbit while also providing state estimates several orders of magnitude lower when compared to the other two laser system's L2D engagements. 
    \item Despite the slow $c_\text{m}$ estimate convergence, the COTS laser system is still capable of reducing the periapsis radius of an ablated debris on the order of meters during a single L2D engagement. While this periapsis radius reduction performance may not be well-suited for deorbiting debris, the COTS laser system achieves highly accurate state estimates; thus, a smaller laser system may be more appropriate for just-in-time collision avoidance operations, where slight perturbations and accurately predicted debris states would be favorable in decreasing the probability of collision with other orbital debris or satellites.
\end{enumerate}

The aforementioned recommendations are made within the context of the experiments designed and analyzed in this work. In each numerical experiment, the ablated debris is considered to be spherical, non-rotating, and a solid homogeneous aluminum rigid body. All debris and laser platforms are initialized in circular orbits. During an L2D engagement, a range-dependent L2D perturbing acceleration is applied to the ablated debris, where a constant momentum coupling coefficient is assumed throughout the entirety of the L2D engagement. Each numerical experiment makes these assumptions, and the results presented in this work are subject to change as assumptions are relaxed; however, the application of the methodology is implemented without loss of generality to more complex L2D engagements when considering initially elliptic debris orbits and various orbital debris characteristics that may require additional dynamical system terms. Relaxing assumptions made on orbital debris characteristics will require an adapted dynamical system and corresponding state vector augmentation contingent on desired parameter estimates during an L2D engagement, but the application of the methodology presented in this work is invariant to changes made to the L2D perturbing acceleration's contribution to the target debris's dynamical system.

\section{Conclusions} \label{sec:Con}
Safely remediating orbital debris via space-based lasers while accounting for uncertainties in remediation operations requires identifying driving factors that contribute to reducing the uncertainty of a given L2D engagement. These factors can reduce or exacerbate errors in both the target debris state and parameter estimates during the remediation campaign. This work provides insights into how target debris periapsis radius lowering, linear system observability, and nonlinear estimation performance vary across changes in encounter geometry and laser system parameters. Two numerical experiments are conducted to investigate how coplanar and out-of-plane L2D engagements are impacted by changes in laser platform and target debris encounter geometries, and a final numerical experiment shows how varying the laser system parameters of a space-based laser affects the L2D engagement for a laser platform engaging a single target debris.

Numerical Experiment 1 shows that a coplanar L2D engagement has a larger total target debris periapsis decrease magnitude for encounter geometries that allow for longer L2D engagements. In these coplanar cases, the length of the L2D engagement increases as the relative altitude between the target debris and the laser platform decreases. When the target debris altitude is greater than the laser platform altitude, then the L2D engagement begins with a minimum periapsis decrease magnitude; conversely, the L2D engagement generally ends with a minimum periapsis decrease magnitude if the target debris altitude is less than that of the laser platform altitude. A general trend for coplanar cases shows that observability and estimation performance increase as the periapsis radius decrease magnitude and rate of change in sampled measurements increase. For out-of-plane L2D engagements, the $k$-means clustering shows how the relative inclination between the laser platform and target debris impacts the overall performance of an L2D engagement. The two clusters' representative case comparison in Numerical Experiment 2 shows that $c_\text{m}$ estimation performance deteriorates for shorter L2D engagements, although the target debris state estimates retain low magnitude $\text{RMSE}_{\text{pos}}$ despite slightly poorer $c_\text{m}$ estimation. Out-of-plane L2D engagements follow the insight gained from the coplanar analysis, where observability and estimation performance increase as the periapsis radius decreases in magnitude and the rate of change in sampled measurements increases. Numerical Experiment 3 shows how varying the laser system parameters affects all aspects of the L2D engagement. An L2D engagement can be ended quickly for a laser system with a sufficiently large time-averaged laser intensity, where the debris is significantly perturbed over a short time interval and quickly exceeds the maximum ablation range. The best state estimation performance for this numerical experiment is given by the COTS system, and the exceptionally low state error is attributed to the significant change in the angle measurements over the L2D engagement.

Based on insights gained from this work, there are several points to investigate in future research. The parameter $c_\text{m}$ governs the L2D perturbing acceleration, but it is one of many uncertain characteristics of the laser-matter interaction during an L2D engagement; therefore, including more uncertainty into the L2D dynamical system via partial knowledge of additional target debris characteristics may give way to new insights when conducting an L2D engagement with a poorly characterized target debris. Additionally, varying laser system parameters show that a low-input-power laser system, such as the COTS laser system, is capable of achieving increased observability and state estimation performance at the detriment of a slowly converging $c_\text{m}$ estimate and low periapsis radius lowering capacity; thus, a future direction for research could lie in investigating the trade-off between increasing the operational range of a laser system while retaining desirable observability and state estimation performance.

\section*{Appendix A: Range-Dependent L2D Perturbing Acceleration Formulation Accounting for Over and Underfill Conditions} \label{app:A}

The range-dependent L2D perturbing acceleration term begins with a description of the laser fluence as a function of the range between the laser platform and target debris. For a given laser system, the pulse energy, $E$, can be defined through Eq.~\eqref{eq:ICAN_E},
\begin{equation}
    E = \pi \left( \frac{d_{\text{L}}}{2} \right)^2 \Psi \label{eq:ICAN_E}
\end{equation}
where the focused spot size at a range $\left\lVert \bm{k} \right\rVert$ is denoted by $d_\text{L}$ \cite{phipps_bonnal_2016}.
The spot size at the range $\left\lVert \bm{k} \right\rVert$ can also be defined as given in Eq.~\eqref{eq:ICAN_spot},
\begin{equation}
    \frac{d_\text{L}}{2} = \frac{c_\text{diff} M^2 \lambda \left\lVert \bm{k} \right\rVert}{2 D} \label{eq:ICAN_spot}
\end{equation}
where the diffraction constant, beam quality factor, and primary mirror diameter are denoted by $c_\text{diff}$, $M^2$, and $D$, respectively.
Substituting Eq.~\eqref{eq:ICAN_E} into Eq.~\eqref{eq:ICAN_spot} and solving for $\Psi$ gives Eq.~\eqref{eq:range_fluence},
\begin{equation}
    \Psi = \frac{4ED^2 T_\text{eff}}{\pi c_{\text{diff}}^2 M^4 \lambda^2 \left\lVert \bm{k} \right\rVert^2} \label{eq:range_fluence}
\end{equation}
which provides a range-dependent laser fluence term as a function of laser system parameters and the range between the laser platform and target debris.

Accounting for a change in the focused spot size of the propagated laser beam requires a specific formulation of the area matrix term included in the L2D perturbing acceleration. The area matrix, $\bm{G}$, needed to describe the surface area of the target debris illuminated by the incident laser beam is defined in Eq.~\eqref{eq:area_matrix_gen},
\begin{subequations}
    \begin{align}
        \bm{G} &= \int \text{d}A \hat{\bm{n}}\hat{\bm{n}} \label{eq:area_matrix_gen} \\
        \hat{\bm{n}} &= \begin{bmatrix}
            \sin(\theta)\cos(\phi) & \sin(\theta)\sin(\phi) & \cos(\theta)
        \end{bmatrix}^\text{T} \label{eq:G_nhat}
    \end{align}
\end{subequations}
where the differential area and the local target debris's surface normal are denoted by $\text{d}A$ and $\hat{\bm{n}}$, respectively \cite{liedahl_2013_pulsed_shape}.
The definition of $\hat{\bm{n}}$ is created in a spherical coordinate system defined in the body reference frame of the spherical target debris, where the origin of the coordinate system is placed at the center of the target debris. The polar axis of this spherical coordinate system extends along the laser beam propagation direction, where $\theta$ and $\phi$ are given as the polar angle and azimuthal angle, respectively. The form of $\hat{\bm{n}}\hat{\bm{n}}$ is a dyadic product, and $\text{d}A$ can be defined as $\text{d}A = \gamma_\text{D}^2 \text{d}\Omega = \gamma_\text{D}^2 \sin(\theta)\text{d}\theta \text{d}\phi$, where $\gamma_\text{d}$ and $\text{d}\Omega$ denote the spherical debris radius and differential solid angle, respectively; furthermore, $\text{d}\Omega$ can be expressed in terms of $\theta$ and the differential polar and azimuthal angles denoted as $\text{d}\theta$ and $\text{d}\phi$, respectively. Substituting this dyadic product and differential area into Eq.~\eqref{eq:area_matrix_gen} with integration limits given as $\theta \in \left[ \pi-\beta,\pi \right]$ and $\phi \in \left[0,2\pi \right]$ gives the area matrix expressed by Eq.~\eqref{eq:area_matrix_int}.
\begin{equation}
    \bm{G} = \gamma_\text{D}^2 \int_{0}^{2\pi} \int_{\pi-\beta}^{\pi} \text{d}\Omega \begin{bmatrix}
        \sin^2\theta \cos^2\phi & \sin^2\theta \sin\phi \cos\phi & \sin\theta \cos\theta \cos\phi \\
        \sin^2\theta \sin\phi \cos\phi & \sin^2\theta \sin^2\phi & \sin\theta \cos\theta \sin\phi \\
        \sin\theta \cos\theta \cos\phi & \sin\theta \cos\theta \sin\phi & \cos^2\theta
    \end{bmatrix} \label{eq:area_matrix_int}
\end{equation}
The illumination angle, $\beta$, given in the lower integration limit for $\theta$, represents the angle between the polar axis and the edge of the illuminated target debris surface. Evaluating Eq.~\eqref{eq:area_matrix_int} shows that off-diagonal terms of the area matrix evaluate to zero, which gives the area matrix in the form given by Eq.~\eqref{eq:area_matrix_int_eval}.
\begin{equation}
    \bm{G} = \gamma_\text{D}^2 \begin{bmatrix}
        \pi \left(\frac{2-3\cos(\beta)+\cos^3(\beta)}{3}\right) & 0 & 0 \\
        0 & \pi \left(\frac{2-3\cos(\beta)+\cos^3(\beta)}{3}\right) & 0 \\
        0 & 0 & 2\pi \left( \frac{1 - \cos^3(\beta)}{3} \right)
    \end{bmatrix} \label{eq:area_matrix_int_eval}
\end{equation}
Expressing $\bar{I}$ in terms of the range-dependent fluence given in Eq.~\eqref{eq:range_fluence} and substituting Eq.~\eqref{eq:area_matrix_int_eval} into Eq.~\eqref{eq:Liedahl_LA_gen}  provides the final form of the range-dependent L2D perturbing acceleration given by Eq.~\eqref{eq:a_L2D_G_eval_main} in Sec.~\ref{subsec:Pulsed_Laser_Ablation}.
\begin{equation}
    \begin{bmatrix}
        a_{\text{L2D,x}}\\
        a_{\text{L2D,y}}\\
        a_{\text{L2D,z}}
    \end{bmatrix} = \begin{bmatrix}
        \frac{4 c_{\text{m}} E D^2 T_{\text{eff}} \upsilon \gamma_\text{D}^2}{c_\text{diff}^2 M^4 \lambda^2 \left\lVert \bm{k} \right\rVert^2 m_\text{D}} \left(\frac{2-3\cos(\beta)+\cos^3(\beta)}{3}\right) \\
        \frac{4 c_{\text{m}} E D^2 T_{\text{eff}} \upsilon \gamma_\text{D}^2}{c_\text{diff}^2 M^4 \lambda^2 \left\lVert \bm{k} \right\rVert^2 m_\text{D}} \left(\frac{2-3\cos(\beta)+\cos^3(\beta)}{3}\right) \\
        \frac{8 c_{\text{m}} E D^2 T_{\text{eff}} \upsilon \gamma_\text{D}^2}{c_\text{diff}^2 M^4 \lambda^2 \left\lVert \bm{k} \right\rVert^2 m_\text{D}}  \left( \frac{1 - \cos^3(\beta)}{3} \right)
    \end{bmatrix} \tag{3}
\end{equation}

\section*{Appendix B: Toric Section Initialization} \label{app:B}

Target debris initialization in the three-dimensional position space is performed with the use of toric sections that are created about the laser platform's circular orbit. Equation~\eqref{eq:torus_xyz} describes the torus created by the maximum laser ablation range as the laser platform moves along a circular orbit,

\begin{equation}
        T(x,y,z) = \left( \sqrt{x^2+y^2}-a_{\text{LP}} \right)^2 + z^2 - L_{\text{max}}^2 \label{eq:torus_xyz}
\end{equation}
where $a_{\text{LP}}$ and $L_{\text{max}}$ are given as the semi-major axis of the laser platform's circular orbit and the maximum distance allowed for an L2D engagement to occur, respectively.

Equation~\eqref{eq:plane_xyz} describes a given target debris's orbital plane through three parametric equations defined in the Cartesian coordinate system \cite{moroni2017toricSec}.

\begin{equation}
        p(\chi,\omega) :
        \begin{cases}
            x = x_Q + \chi\sin(\Omega_{d}) - \omega\cos(\Omega_{d})\sin(i_{d}) \\ 
            y = y_Q - \chi\cos(\Omega_{d}) - \omega\sin(\Omega_{d})\sin(i_{d}) \\ 
            z = z_Q + \omega\cos(i_{d})
        \end{cases}, ~\text{where}\begin{cases}
            x_Q = \rho_{\text{tor}}\cos(\Omega_{d})\cos(i_{d})\\
            y_Q = \rho_{\text{tor}}\sin(\Omega_{d})\cos(i_{d})\\
            z_Q = \rho_{\text{tor}}\sin(i_{d})
        \end{cases}
        \label{eq:plane_xyz}
\end{equation}
The two Cartesian coordinates defining the two-dimensional position space on the target debris orbital plane are denoted as $\chi$ and $\omega$, and the distance of this plane's origin with respect to the origin of the Earth-centered Inertial (ECI) frame is denoted as $\rho_{\text{tor}}$. For this work, it is assumed that $\rho_{\text{tor}}$ is equal to zero; therefore, the component-wise displacement of the orbital plane origin with respect to the ECI frame origin, $x_Q$, $y_Q$, $z_Q$, is equal to zero. Additionally, the target debris's right ascension of the ascending node and inclination are denoted as $\Omega_d$ and $i_d$, respectively.

Equation~\eqref{eq:tor_sec_in_int_plane3D} describes the toric section lying on the orbital plane created by the intersection of the torus formed about the laser platform orbit and the target debris's orbital plane.
\begin{equation}
        \tau(\omega) : \begin{cases}
            x = x_Q + \chi(\omega)\sin(\Omega_{d})-\omega\cos(\Omega_{d})\sin(i_{d}) \\
            y = y_Q - \chi(\omega)\cos(\Omega_{d})-\omega\sin(\Omega_{d})\sin(i_{d}) \\
            z = z_Q + \omega\cos(i_{d})
        \end{cases}
        \label{eq:tor_sec_in_int_plane3D}
\end{equation}
where,

\begin{equation*}
    \label{eq:tor_tw_to_xyz}
    \chi(\omega) = \pm\sqrt{-(\rho_{\text{tor}}\cos({i_{d}})-\omega\sin(i_{d}))^2 + \left(a_{\text{LP}}\pm\sqrt{L_{\text{max}}^2-(\omega\cos(i_{d})+\rho_{\text{tor}}\sin(i_{d}))^2}\right)^2}
\end{equation*}

\begin{figure}[H]
    \centering
    \includegraphics[width=.5\linewidth]{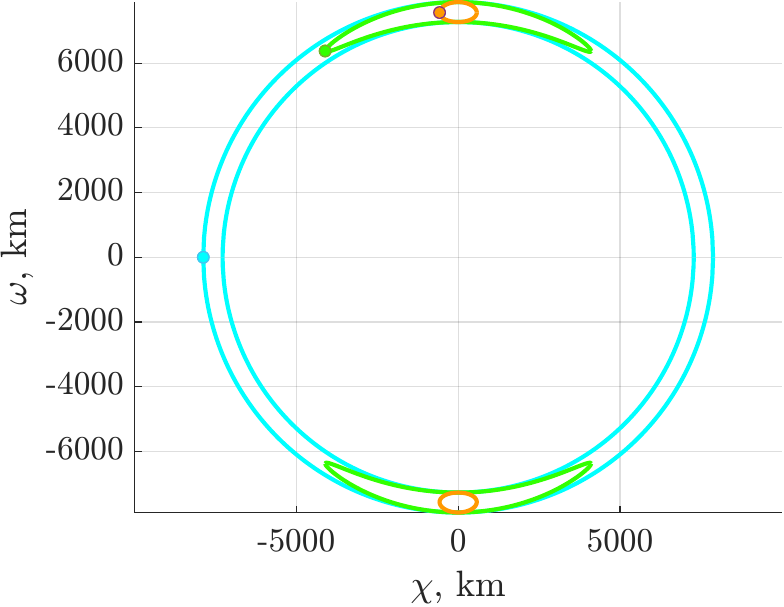}
    \caption{Example target debris positions given on toric sections, where initial target debris positions are given as scatter points.}
    \label{fig:tor_orb_plane}
\end{figure}

\begin{figure}[H]
    \centering
    \includegraphics[width=.55\linewidth]{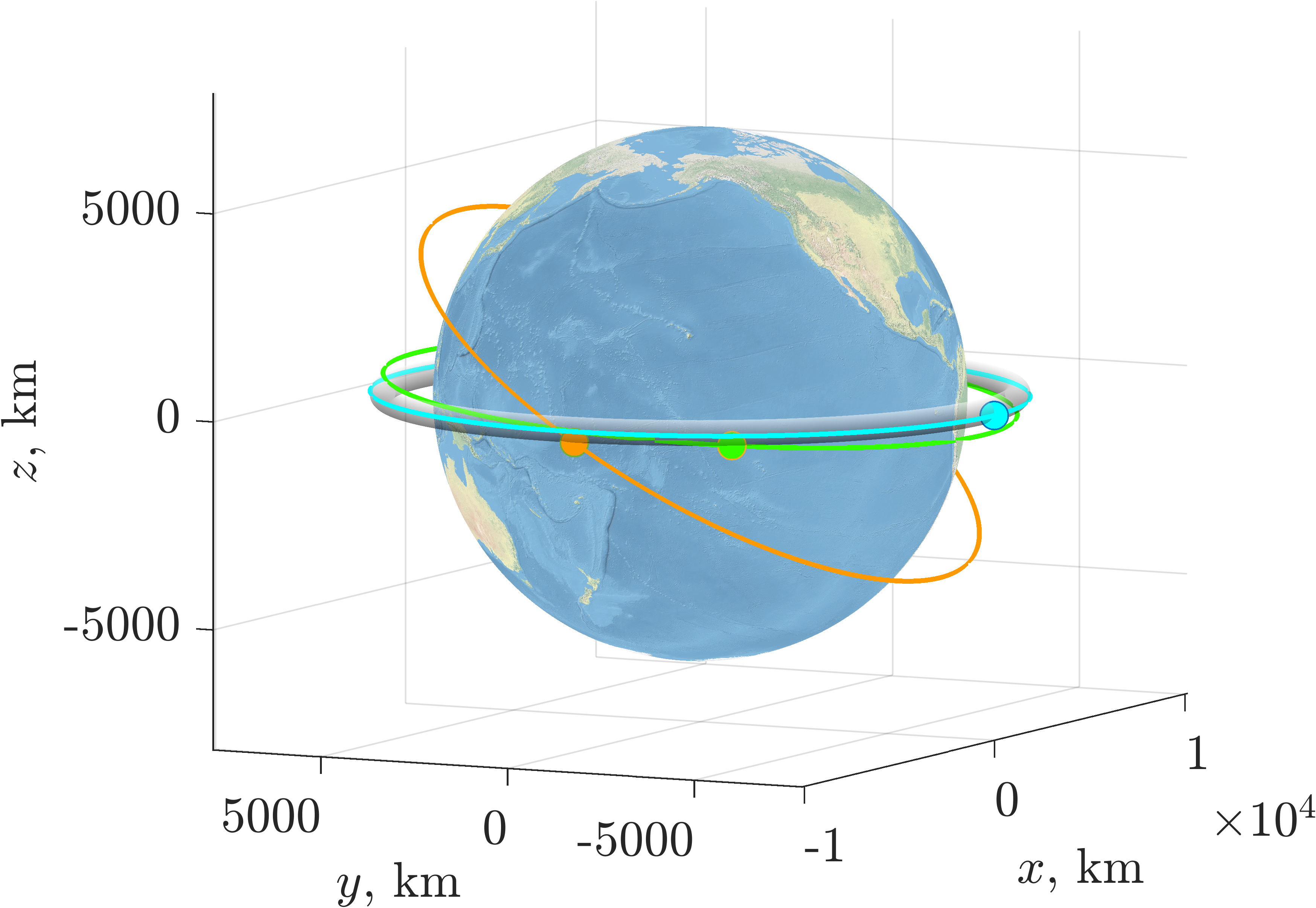}
    \caption{Example target debris orbits given in the ECI frame, where colored scatter points are the initial position for the same color propagated orbit.}
    \label{fig:tor_ECI}
\end{figure}

In Numerical Experiment 2, target debris inclinations are logarithmically spaced from \SI{0.01}{deg} to \SI{89}{deg} with 10 inclination bins being created. For each inclination bin considered, the initial $\omega$ position is linearly spaced from \SI{0}{km} to \SI{7885.9}{km}. The positions on the orbital plane in terms of $(\chi,\omega)$ are then converted to positions in the ECI frame according to Eq.~\eqref{eq:tor_sec_in_int_plane3D}. Toric sections without distinct multiple regions on the orbital plane had duplicate initializations removed, which resulted in a total of $96$ initialized target debris positions. Three distinct target debris generation cases using the toric sections are shown in Figs.~\ref{fig:tor_orb_plane} and \ref{fig:tor_ECI}, where the cyan, green, and orange colors represent varying orbital inclinations. Initialized target debris positions are then propagated forward in time to the timestep where each debris would enter the laser ablation maximum range at the next timestep. This resulted in two octants of the sphere defining possible L2D engagements to be populated with target debris having lower and higher magnitude altitudes compared to the laser platform initialized on the negative $x$-axis with a circular equatorial orbit and a semi-major axis of \SI{7578.14}{km}. The propagated target debris state vectors are then treated as the initial state vectors used for each target debris and laser platform pair in Numerical Experiment 2.

\clearpage
\newpage
\section*{Acknowledgments}
This work was supported by an Early Career Faculty grant from NASA's Space Technology Research Grants Program under award No. 80NSSC23K1499.

\bibliography{references}

\end{document}